\documentclass[prd,twocolumn,preprintnumbers,superscriptaddress,nofootinbib]{revtex4}
\usepackage[a4paper, hdivide={1.91cm,,1.165cm}, vdivide={1.83cm,,2.6cm}]{geometry}

\usepackage{amstext,amssymb}
\usepackage{amsmath,lipsum}
\usepackage{graphicx}
\usepackage{subcaption}
\usepackage[toc,page]{appendix}
\usepackage[hyperfootnotes=false]{hyperref}
    \usepackage{hyperref}
\hypersetup{ 
	setpagesize=false,
	bookmarksnumbered=true,
	colorlinks=true,
	linkcolor=blue,
	citecolor=red,
	hypertexnames=true
}
\usepackage{xspace}
\usepackage{color}
\usepackage{caption}
\usepackage{subcaption}
\usepackage{float}
\usepackage{units}
\usepackage{slashed} % feynman slash notation via \slashed{p}
\usepackage{braket}

\def \vec#1{{\boldsymbol{#1}}}

\begin{document}

\title{Geometrical Interpretation of Neutrino Oscillation with decay}
% %\preprint{ULB-TH/15-10}
\author{Rajrupa \surname{Banerjee}}
\email{rajrupab@iitbhilai.ac.in}
% \affiliation{Physical Research Laboratory, Navarangpura, Ahmedabad, India}
\affiliation{Department of Physics, Indian Institute of Technology (IIT) Bhilai, Bhilai 491001, India}
\author{Kiran \surname{Sharma}}
\email{kirans@iitbhilai.ac.in}
\affiliation{Department of Physics, Indian Institute of Technology (IIT) Bhilai, Bhilai 491001, India}
\affiliation{Institut für Astroteilchenphysik, Karlsruher Institut für Technologie(KIT), Hermann-von-Helmholtz-Platz 1, 76344, Eggenstein-Leopoldshafen, Germany.}

\author{Sudhanwa \surname{Patra}}
\email{sudhanwa@iitbhilai.ac.in}
\affiliation{Department of Physics, Indian Institute of Technology (IIT) Bhilai, Bhilai 491001, India}

\author{Prasanta K. \surname{Panigrahi}}
 \email{panigrahi.iiser@gmail.com}
\affiliation{Indian Institute of Science Education and Research Kolkata, 741246, WB, India}

%

%%%%%%%%%%%%%%%%%%%%%
\begin{abstract}
%
%The W-mass anomaly has not been established, but a massive proliferation of articles on the subject showed the rich potential of such events.
The geometrical representation of two-flavor neutrino oscillation represents the neutrino's flavor eigenstate as a magnetic moment-like vector that evolves around a magnetic field-like vector that depicts the Hamiltonian of the system. In the present work, we demonstrate the geometrical interpretation of neutrino in a vacuum in the presence of decay, which transforms this circular trajectory of neutrino into a helical track that effectively makes the neutrino system mimic a classical damped driven oscillator. We show that in the absence of the phase factor $\xi$ in the decay Hamiltonian, the neutrino exactly behaves like the system of nuclear magnetic resonance(NMR); however, the inclusion of the phase part introduces a $CP$ violation, which makes the system deviate from NMR.  Finally, we make a qualitative discussion on under-damped, critically-damped, and over-damped scenarios geometrically by three different diagrams. In the end, we make a comparative study of geometrical picturization in vacuum, matter, and decay, which extrapolates the understanding of the geometrical representation of neutrino oscillation in a more straightforward way.

\end{abstract}
%%%%%%%%%%%%%%%%%%%%
\maketitle 
\section{Introduction} 
\label{sec:intro}
%Neutrinos are one of the most exciting particles in nature. 
%\cite{Mohapatra:2005wg}.
%Generically, they are categorized as neutral fermions %\cite{Halzen:1984mc}\cite{Bilenky:2004xm}\cite{Abazajian:2012ys}
%. But their exact fundamental nature is still unknown \cite{Bilenky:1987ty}\cite{Bilenky:2020wjn}. 

The experimental shreds of evidence of neutrino oscillation have proven the existence of tiny but nonzero masses of neutrino, which leads to a new era of exploration of fundamental physics~\cite{SNO:2002tuh, Super-Kamiokande:2002ujc, IceCube:2017lak, DayaBay:2012fng, MINOS:2011amj}. But for a better understanding of neutrino oscillation visually, the geometrical interpretation of neutrino oscillation had been proposed where neutrino oscillation was described as the rotation of a unit vector depicting the flavor state of the neutrinos around a magnetic field-like vector denoted by the Hamiltonian of the neutrino system \cite{Giunti:2007ry, Kim:1987bv, Smirnov_1986jht}. Analogously, this picturization has been compared with the precession of the magnetic moment vector in the presence of an external magnetic field. The projection of the flavor state $\ket{\nu_{\alpha}(t)}$ at time t, over the mass basis $\ket{\nu_{i}}$ is connected to the amplitude of oscillation \cite{Giunti:2007ry}.

 Although the geometrical interpretation makes it easier to understand the neutrino oscillation and realize it pictorially, such representation is not unique. Depending on the different choices of basis, some mathematical approaches have come up with a heuristic overview of the flavor state oscillation. For example, in \textit{Giunti. et al.}~\cite{Giunti:2007ry} an orthogonal basis $e_{F}^{i}$, $e_{M}^{i}$ describing the flavor and the mass basis of the neutrino has been investigated, where \textit{i} stands for 1,2,3. The flavor state $\nu_{e}$ revolves around the Hamiltonian vector $\Vec{B}$ with an angle $2\theta$. A similar kind of picturization has been done in \textit{Smirnov. et al}~\cite{Smirnov_1986jht} with the choices of basis with axes $\nu_{1}$, $Re\nu_{1}$, $Im\nu_{2}$ to describe the two generation neutrino oscillation. On the other hand, \textit{Kim. et al.}~\cite{Kim:1987bv, Sen:2018put} discussed the geometrical interpretation of flavor oscillation of neutrino in three-dimensional Euclidean space deduced from the two-valued representation of the flavor space. They implemented their interpretation to study the MSW effect in adiabatic and non-adiabatic scenarios.

  Oscillation experiments \cite{JUNO:2021ydg, Ghoshal:2020hyo, Gonzalez-Garcia:2008mgl} and extensive literature \cite{Bahcall:1972my, Lindner:2001fx, Blasone:2020qbo} ambiguously prove that neutrinos possibly decay to lighter invisible states. The literature proclaims an extensive study of the analytical approach to neutrino oscillation in the presence of decay \cite{Chattopadhyay:2021eba, Dixit:2022izn}. However, due to the non-Hermiticity of decay-Hamiltonian, it is challenging to analyze the probability of oscillation analytically, making it harder to visualize the picture of the damped neutrino oscillation \cite{Chattopadhyay:2021eba, Goswami:2023hux}. Due to this, a geometrical approach to understanding the oscillation in the presence of decay Hamiltonian has not been proposed yet.

 In the present work, we address the above adversity of the non-hermitian nature of decay Hamiltonian by a simple mathematical approach analogous to the NMR system and illustrate the neutrino oscillation geometrically in the presence of decay Hamiltonian. The motivation for adopting such a procedure is to consider neutrino as an open quantum system with a similar viewpoint to an NMR system where the external magnetic field acts as a beam splitter; in the case of neutrino, the unitary transformation matrix performs the same \cite{KumarJha:2020pke, Blasone:2010ta, Blasone:2019chv, Blasone:2020qbo}. The result of the damped oscillation of neutrino also goes hand in hand with the NMR. However, introducing the phase factor $\xi$ with the non-Harmitian decay Hamiltonian the introduces $CP$ violation in two flavor neutrino oscillation even in absence of the Majorana phase which modifies the dynamic of the flavor state that introduces a significant difference from the NMR. 
 For a simplistic overview, we restrict ourselves to only two-flavor neutrino oscillation.

 The construction of the paper is as follows. We start our work with the general theoretical framework(Section II) for the neutrino oscillation, where we define the two mass eigen states $\ket{\nu_{1}}$ and $\ket{\nu_{2}}$ as a two-level system. We also introduce the decay Hamiltonian along with it. Next, we follow the density matrix formalism to define the neutrino as an open quantum system in the presence of decay. To picture the oscillation vectorially, we vectorized the operator using the bra-flipper operator. We use the Pauli Matrix basis to represent the vectorized operator and construct the dynamical evolution equation. In the result and discussion session(section III), we evaluate the probabilities and equation of motion. We explain our result with two different schemes. First, we approximated the decay parameters to be zero for the ideal case. In the next, we derive the equation of motion, including the decay parameters. In this section, we also explain the result graphically and present the same using geometrical schematics. In the successive section(section IV), we compare geometrical interpretation in different bases following \cite{Kim:1987bv, Sen:2018put} with the present work. We also briefly discuss the matter effect and illuminate the geometrical interpretation of neutrino oscillation in matter.
The paper concludes with an overall summary of the total work and sheds light on some of the other perspectives of the work for future reference.

%%%%%%%%%%%%%%%%%%%%%%%%%%%%%%%%%%%%%%%%%%%%%%%%%%%%%%%%%%%%%%%%%%%%
%%%%%%%%%%%%%%%%%%%%%%%%%%%%%%%%%%%%%%%%%%%%%%%%%%%%%%%%%%%%%%%%%%%%
\section{Theoretical Framework}
\label{sec:Theory}
Consider an ultra-relativistic neutrino (left-handed) with flavor $\alpha$ with $\alpha=e,\mu,\tau$ and momentum $\Vec{p}$. The present theoretical framework extends to considering the following two eigenstates of neutrinos, i.e., mass eigenstate, which is the propagating eigenstate, and flavor eigenstate, which is the interacting eigenstate. These two eigenstates are expressed by the two state vectors  $\ket{\nu_{\alpha}}$ and $\ket{\nu_{i}}$ which are related by the unitary transformation transformation.
\begin{equation}
    \ket{\nu_{\alpha}}=\sum_{k}U^{*}_{\alpha k}\ket{\nu_{k}}
\end{equation}\label{eq:1}
Where $\ket{\nu_{\alpha}}$ is the flavor state of the neutrino, and $\ket{\nu_{k}}$ is the mass eigenstate of the neutrino. 
As we deal only with the two flavors, no additional $CP$ phase part is needed. 
Hence, the unitary transformation $U^{*}$ will be replaced by the orthogonal transformation $\mathcal{O}^{T}$.   where,
\begin{equation}
    \mathcal{O}=\begin{pmatrix}
        \cos{\theta} & \sin{\theta}\\
        -\sin{\theta} & \cos{\theta}
    \end{pmatrix}
\end{equation}
In this work, we consider only Dirac-type neutrino. It is to be mentioned here that for Majorana type of neutrino $CP$ violation phase needed to be considered even for the two-flavor neutrino oscillation as well\cite{Dixit:2022izn}, i.e., $\ket{\nu_{\alpha}}=\mathcal{O}U_{ph}\ket{\nu_{i}}$, where $U_{ph}=\begin{pmatrix}
    1 & 0\\
    0 & e^{i\phi}\\
\end{pmatrix}$.
Associated with any physical system, there is a complex Hilbert space known as the system's state space. In Hilbert space, the state vector entirely describes the system's state. 
%A state vector, a unit vector in the Hilbert space, describes the system's state entirely. 
These state vectors of a finite Hilbert space correspond to the possible physical condition of the system, and they are the pure state.

 Considering the fact that a column matrix gives any state vector in Hilbert space, we consider the mass eigenstates of the neutrino as a two-level system with state $\ket{\nu_{1}}=
 \begin{pmatrix}
    1\\
    0\\
\end{pmatrix}
$ and state $\ket{\nu_{2}}=\begin{pmatrix}
    0\\
    1\\
\end{pmatrix}$.
In the present work, we restrict ourselves to the two-flavor oscillation only. In Hilbert space any operator is given by a square matrix, $\Hat{A}=\begin{pmatrix}
    \bra{1}\Hat{A}\ket{1} & \bra{1}\Hat{A}\ket{2}\\
    \bra{2}\Hat{A}\ket{1} & \bra{2}\Hat{A}\ket{2}\\
\end{pmatrix}$
With diagonal elements as the eigenvalues and the off-diagonal elements as the transition elements between the two levels. In resemblance with the neutrino system states $\ket{1}$ and $\ket{2}$ replace by the mass eigenstates $\ket{\nu_{1}}$ and $\ket{\nu_{2}}$.
To start with, consider two flavor neutrino states as $\psi = \big(\nu_e\, \quad \nu_\mu \big)^T$ and the corresponding effective Hamiltonian in vacuum is expressed in the limit of ultra-relativistic energy approximation $E_i \simeq p+m_i^2/2p \approx E + m_i^2/2E$ as,
\begin{equation}	
\mathcal{H}_{0}=\left[E \left(
	\begin{array}{cc}
	 1 & 0 \\
	 0 & 1
	\end{array}
	\right)+\frac{1}{2E}\left(
	\begin{array}{cc}
	 m^2_1 & 0 \\
	 0 & m^2_2
	\end{array}
	\right)\right]
\label{eq:schrod2}
\end{equation}
It is well known that the term proportional to identity has no effect on neutrino oscillation and hence can be dropped from now on. The neutrino mass eigenstates are related to the corresponding flavor eigenstates by a unitary mixing matrix. Thus, the vacuum part of effective Hamiltonian in flavor basis is read as,
 \begin{eqnarray}
\hspace*{-0.5cm} \mathcal{H}_{\rm F} \hspace*{-0.2cm}&=& \hspace*{-0.2cm}\frac{1}{2 E}
  \left[
	\begin{array}{cc}
	 m^2_1 \cos^2\theta+m^2_2 \sin^2\theta & -\cos\theta \sin\theta \Delta m^2_{} \\
	 -\cos\theta \sin\theta \Delta m^2_{} & m^2_1 \sin^2\theta_+m^2_2 \cos^2\theta
	\end{array}
	\right]
  \nonumber \\
  &=&\frac{\Delta m^2_{}}{4E}  \begin{pmatrix} 
    -  \cos2\theta & \sin2\theta \\
      \sin2\theta &  \cos2\theta \\
 \end{pmatrix} \big]
 \label{eq:H-vac}
 \end{eqnarray}
with $\Delta m^2_{} = m^2_2-m^2_1$ is the mass square difference between the two mass eigenvalues of the neutrinos.
%For the two-flavor neutrino oscillation in vacuum, the Hamiltonian operator is given by,\cite{Dixit:2022izn}
%\begin{equation}\label{eq:4}
 %   \mathcal{H}_{0}=\begin{pmatrix}
  %      \frac{m^{2}_{1}}{2E} & 0\\
   %     0 & \frac{m^{2}_{2}}{2E}\\
    %\end{pmatrix}
%\end{equation}
From the eigenvalue equation for mass eigenstate have the eigenvalues
\begin{equation}
\vspace{-0.3cm}
\mathcal{H}_{0}\ket{\nu_{k}}=E_{k}\ket{\nu_{k}}
\end{equation}
 $E_{1}=\frac{m^{2}_{1}}{2E}$ and $E_{2}=\frac{m^{2}_{2}}{2E}$.
Here we consider the same energy approach for the ultra-relativistic process $p_{i}\sim E+\frac{m^{2}_{i}}{2E}$ so that the oscillation phase reduces to $\Delta\phi=-\Delta p\cdot L\sim \frac{\Delta m^{2}}{2E}L$ \cite{Akhmedov:2019iyt}.
It is sufficient to notice no transition between the two energy levels in vacuum oscillation.
In Schrodinger's picture, any neutrino state follows the evolution equation of the initial flavor $\alpha$ as
\begin{equation}
    i\frac{d}{dt}\ket{\nu_{\alpha}}=\mathcal{H}_{F}\ket{\nu_{\alpha}}
\end{equation}
Where $\mathcal{H}_{F}$ is the Hamiltonian of the flavor state. For simplified of the mathematical description we take $\frac{m^{2}_{1}}{2E}=a_{1}$ and $\frac{m^{2}_{2}}{2E}=a_{2}$.

From the experimental evidence of neutrino oscillations, it has already been proven that neutrino has a finite mass. All the particles having finite rest mass have some finite decay width. The decay probability increases with the mass of the particles \cite{Halzen:1984mc, Baerwald:2012kc}. The decay probability becomes very small for a very low-mass particle, and we get quasi-stationary states $i.e.$, the particles are localized for a long time. Without decay, each flavor eigenstate is the coherent superposition of mass eigenstates.
 The Hermitian nature of the Hamiltonian gives rise to real eigenvalues. Nevertheless, the non-Hermitian attribute of the decay Hamiltonian generates discrete complex eigenvalues. The Decay Hamiltonian for neutrino is given by\cite{Dixit:2022izn, Borsten:2008wd, Berryman:2014yoa, Chattopadhyay:2021eba}
\begin{equation}
\Gamma/2= \begin{pmatrix} 
b_{1} & \frac{1}{2}\eta e^{i\zeta} \\
\frac{1}{2}\eta e^{-i\zeta} & b_{2}
 \end{pmatrix}
 \end{equation}
Incorporating the decay Hamiltonian with the standard Hamiltonian term, we get the effective Hamiltonian $\mathcal{H}_{m}$ in mass basis,
\begin{equation}\label{eq:7}
\begin{split}
    \mathcal{H}_{m} & =\mathcal{H}_{0}-i\Gamma/2\\
                 & ={\begin{pmatrix}  
                    a_{1} & 0 \\       
                    0 & a_{2}          
                    \end{pmatrix}}-
                    {\begin{pmatrix} 
                    ib_{1} & \frac{i}{2}\eta e^{i\zeta} \\
                    \frac{i}{2}\eta e^{-i\zeta} & ib_{2}
                    \end{pmatrix}}
\end{split}
\end{equation}
$\mathcal{H}_{0}$ is the vacuum Hamiltonian, associated with the mass square terms and energy. Considering a finite decay width of the single neutrino system, a decay Hamiltonian is also associated with the vacuum Hamiltonian given by the $\Gamma/2$ matrix. As mentioned in \cite{Dixit:2022izn} a bound of $\tau_{\nu} \geq 5.7\times 10^5
s (m_{\nu} /eV)$ is derived from the neutrino data of Supernova 1987A.
which leads to $\Gamma_{\nu} \equiv b\approx 10^{-21} eV$ for a neutrino of mass 1 eV. Also, in \cite{Soni:2023njf} suggested an approximate value of $b$. Although the literature suggests a rough estimation of the value of the decay parameter, it converges towards the ultrahigh-energy neutrino from the astrophysical sources \cite{Dixit:2022izn}. Here, $b_{1},b_{2}$ and $\eta$ are real parameters depicting the cause of the decay of mass eigenstate. Since there is no established numerical value of decay parameters, the motivation of the work is to visualize the decay geometrically and following the defined value of the parameters and probability in\cite{Dixit:2022izn,Chattopadhyay:2021eba, Soni:2023njf, Shafaq:2021lju} plot the appearance and disappearance probabilities in the presence of decay. For the presence work, we restrict ourselves only to the invisible neutrino decay for which we approximate $\xi=0$\cite {Shafaq:2021lju}. Also, for analytical simplicity, we consider $b_{1}=b_{2}=b$ \cite{Dixit:2022izn}.
%Any physical system is divided into two parts 
%\begin{itemize}
 %   \item The fast variables that change rapidly in time
  %  \item Slow variable that can be treated as a fixed or slowly varying parameter.
%\end{itemize}
%The electronic variables of a molecule are an example of fast variables, and the nuclear coordinates are the slow variables. In this case, the motivation is to investigate the moving variable effect and modify the dynamical equation of the fast-moving variables.
\\ The claim is that the Hamiltonian of the system doesn't change with time. The decaying system makes the standard Hamiltonian of the system, $\mathcal{H}_{0}$, non-commutative with the decay Hamiltonian, $\Gamma/2$, as a non-hermitian off-diagonal element $\eta$ is associated with the decay Hamiltonian. However, in spite of the anti-Hermitian nature of the diagonal element $b$ of $\Gamma/2$, it commutes with $\mathcal{H}_{0}$. This is so because, in the absence of the off-diagonal element, the decay basis matches with the mass basis Hence, the information about the eigenstates of such a decaying is not explicitly deterministic. Hence, the intuition is to find a basis for this modified Hamiltonian for a system of decaying neutrino oscillation and to interpret the system's dynamics geometrically.
%For simplification we consider the $b_{1}=b_{2}=b$ \cite{Dixit:2022izn}. Also, for the invisible neutrino decay in vacuum, we take $\xi=0$ \cite{Shafaq:2021lju}\cite{Ohlsson:2000mj},.The claim is that the Hamiltonian of the system doesn't change with time. But it is also associated with a damping matrix, which is noncommutative with the Hamiltonian of the system. Hence, the information about the eigenstates of such a decaying is not explicitly deterministic. The intuition is to find a basis for this modified Hamiltonian for a system of decaying neutrino oscillation and to interpret the system's dynamics geometrically. The simplified Hamiltonian is given by,
We represent the Hamiltonian for the decaying system of the neutrino in the mass basis as $\mathcal{H}_{m}$, which comprise both $\mathcal{H}_{0}$ and $\Gamma/2$, Figuratively, 
\begin{equation}\label{eq:8}
\begin{split}
    \mathcal{H}_{m} & =\begin{pmatrix}
        a_{1} & 0\\
        0 & a_{2}\\
    \end{pmatrix}-\frac{i}{2}\begin{pmatrix}
        b & 0\\
        0 & b\\
    \end{pmatrix}-\frac{i}{2}\begin{pmatrix}
        0 & \eta/2\\
        \eta/2 & 0\\
    \end{pmatrix}
\end{split}
\end{equation}
\
%The Diagonal matrix associated with $b$ of the decay Hamiltonian coincides with the mass basis. It can be associated with the mass basis of the vacuum Hamiltonian. It is also to be mentioned that the diagonal matrix of the decay Hamiltonian commutes with the vacuum Hamiltonian. But the off-diagonal matrix is noncommutative. Hence, it is associated with the transition of the mass eigenstate, i.e., decay of the mass eigenstate. Simplifying the Hamiltonian,
%\begin{equation}
%\begin{split}
 %   \mathcal{H}_{m} & =\begin{pmatrix}
  %      z_{1} & 0\\
   %     0 & z_{2}\\
    %\end{pmatrix}-\frac{i}{2}\begin{pmatrix}
        %0 & \eta/2\\
        %\eta/2 & 0\\
    %\end{pmatrix}\\
%\end{split}
%\end{equation}
%Where $z_{1}=a_{1}-ib/2$ and $z_{2}=a_{2}-ib/2$ In the simplified form, the Hamiltonian in mass basis incorporating decay is,
We redefine eq.(\ref{eq:9}) as $\frac{(a_{1}+a_{2})}{2}\sigma_{0}+\frac{(a_{1}-a_{2})}{2}\sigma_{3}-ib\sigma_{0}-i\frac{\eta}{4}\sigma_{1}$, \cite{Dixit:2022izn}. Explitice mathematical form of this description is, 
\begin{equation}
\label{eq:9}
\begin{split}
    \mathcal{H}_{m} & =\omega\sigma_{3}+(-ib)\sigma_{0}+(-i\eta)/2\sigma_{1}
\end{split}
\end{equation}
The term, $\omega=\frac{\Delta m^{2}}{4E}$, incorporates the mass square differences of the neutrinos, which is the most essential component for the phenomenon of oscillation. We neglect here the term $\frac{(a_{1}+a_{2})}{2}$ in $\sigma_{0}$ as it has no explicit physical significance in flavor-changing neutrino oscillation. This 
decomposition makes it possible to represent the Hamiltonian structurally as the inner product of a vector with the Pauli matrices. To address this affirmation, we introduce a vector $\mathcal{B}$, which denotes the vector representing $\mathcal{H}_{m}.$
This Hamiltonian is now analogous to the inner product of two vectors. Now define a vector $\mathcal{B}$, depicting the Hamiltonian of the system. The above expression is analogous to the inner product of the vectorized Hamiltonian, $\mathcal{B}$ and the Pauli matrices, which gives as follows,
\begin{equation}
\mathcal{H}_{m}=\mathcal{B}\cdot\mathbf{\sigma}
\end{equation}
However, the ambiguity arises in defining the basis. Instead of relating Cartesian coordinates to define the Pauli matrix, we use the mass eigenstates, $\ket{\nu_{1}}$ and $\ket{\nu_{2}}$ to expound the Pauli matrices. This characterization is feasible for the present framework as we consider mass and flavor basis. In this context, we refurbish the Pauli matrices,
\begin{equation}\label{eq:12}
\begin{split}
    \sigma_{x} &=\ket{\nu_{1}}\bra{\nu_{2}}+\ket{\nu_{2}}\bra{\nu_{1}}\\
    \sigma_{y} &=i\ket{\nu_{1}}\bra{\nu_{2}}-i\ket{\nu_{2}}\bra{\nu_{1}}\\
    \sigma_{z} &=\ket{\nu_{2}}\bra{\nu_{2}}-\ket{\nu_{1}}\bra{\nu_{1}}\\
    I_{2} & =\ket{\nu_{1}}\bra{\nu_{1}}+\ket{\nu_{2}}\bra{\nu_{2}}\\
\end{split}
\end{equation}
Eq.(\ref{eq:12}) suggests the operator representation of the Pauli matrices. 
Consequently, this definition motivates us to represent a new basis for the Pauli matrix. However, two points need to be highlighted. Firstly, the dimension of this space, i.e., $\ket{\nu_{i}}\bra{\nu_{j}}$ will differ from the previously defined state space $\ket{\nu_{i}}$. Secondly, a vector is represented by a column matrix, whereas a square matrix represents an operator. When we modify the Hamiltonian operator formalism to vector representation, the square matrix must also be converted to the column matrix. This is possible using a bra-flipper operator $\mho$, \cite{Gyamfi_2020},
\begin{equation*}
\mho[\ket{\nu_{i}}\bra{\nu_{j}}]=\ket{\nu_{i}}\otimes\ket{\nu_{j}}^{*}\equiv|\nu_{i},\nu_{j}\rangle\rangle
\end{equation*}
 % We can create a Hilbert space of density matrices by defining a scalar product. This is clear for finite systems because, in this case, the scalar product and the Hilbert space are the same things. It is also to be true for the infinite space.
 This allows us to define a linear space of matrices, converting the matrices effectively into vectors \cite{Gyamfi_2020, Manzano:2020abc}. But the representation also has an extra complex factor $i$, making it difficult to interpret the $\mathcal{B}$ vectorially. 
 Here, we initiate the concept of quaternion \cite{Wharton:2014fdg}.
 The quaternion is the three-dimensional projection of a four-dimensional complex hyperspace, similar to stereographic projection in complex analysis \cite{Rotelli:1988fc}. It is noteworthy that Pauli matrices form the quaternion groups \cite{Shirokov_2018oiu}. Hence, it is physically more relevant to use the Pauli matrices as the basis for this vectorization of the operator, which inherently carries the complex factor within. Observing such fact in Pauli basis
 , the complex $i$ factor is nothing but a phase factor $e^{-i\pi/2}$ which denotes the rotation of the axis \cite{Bohm_1986uoi}. This implies that the inner product with Pauli matrices leads to the vectorization of the operator. Following this, the Hamiltonian modifies to,
 %The Hamiltonian is represented on this basis as,
 \begin{equation}\label{eq:13}
     \mathbf{B}=\omega\Hat{\sigma}_{3}+(-ib)\sigma_{0}+e^{-i\pi/2}\eta\Hat{\sigma}_{1}
 \end{equation}
 $\Hat{\sigma}$ is the correspond
 $\mathbf{B}$ denotes the vectorized Hamiltonian in mass basis. Therefore, moving it in the flavor basis changes the basis vectors, i.e., the Pauli operator. The rotational transformation of the Pauli basis which is $\sigma_{0}^{M} =\sigma_{0}^{F}$,$ \sigma_{1}^{M} =\sin{2\theta}\sigma_{3}^{F}+\cos{2\theta}\sigma_{1}^{F}$, $\sigma_{2}^{M} =\sigma_{2}^{F}$, $\sigma_{3}^{M} =\cos{2\theta}\sigma_{3}^{F}-\sin{2\theta}\sigma_{1}^{F}$. Defining this, we next move to analyze the flavor state and its evolution on this new basis.
 \\ Now consider the neutrino is created in the flavor eigen state as electron neutrino in Hilbert space $\mathcal{H}_{1}$. We have already mentioned that each flavor eigenstate of neutrino does not have any definitive eigenstate. Instead, it is the linear superposition of the two different mass eigenstates.  The dynamical propagation of the mass basis acquires an amount of flavor basis during propagation, due to which a certain amount of mass also gets mixed associated with each flavor state. 
 %During propagation, the mass, and the flavor eigenstates oscillate amongst each other, for which each mass basis acquires some properties of the different flavor bases and vice versa. 
 More details can be followed from \cite{smirnov2019mikheyevsmirnovwolfenstein, Smirnov:2003da}. 
 As we have already mentioned, each flavor state of neutrino can be thought of as a two-level system consisting of two orthonormal mass eigenstates, $\ket{\nu_{1}}=\ket{0}$ and  $\ket{\nu_{2}}=\ket{1}$, defined over the Pauli bases.
% The explicit matrix form of these bases has already been mentioned. 
 For the present discussion on the two-flavor oscillation, we restrict the superposition to the two mass eigenstates only, i.e., $\ket{\nu_{\alpha}}=\cos{\theta}\ket{0}+\sin{\theta}\ket{1}$. Hence, each flavor state of neutrino is an entangled state of the mass eigenstate \cite{KumarJha:2020pke,  Blasone:2019chv}.
 From the basic postulates of the quantum mechanics corresponding to every system's physical state, a state vector is associated with it. The state vector represents the pure states, which specify all the known physical information about the system. This is an ideal case. In most cases, the system is not in a pure state. Most generally, we may know that a quantum system can be in one state of a set $\{\ket{\nu_{\alpha}}\}$ with probabilities $w_{i}$. In this case, the mathematical tool that describes our knowledge of the system is the density operator (or density matrix). Due to the oscillatory property of the neutrinos, instead of using a state vector for the flavor state, we use the density operator $\Hat{\rho}_{\alpha}$ where $\Hat{\rho}_{\alpha}\in\mathcal{H}_{1}$ \cite{Shafaq:2021lju}.
%Since the flavor state of neutrino is oscillating, and it is the linear superposition of the two mass eigenstates while propagating, an admixture corresponds to the mass eigenstate at the given flavor state. Due to this fact, the mass eigenstate of a particular flavor is unknown. Moreover, depending on the weight factor of mixing each mass eigenstate, there is a transition between the two flavor states. 
Consequently, instead of investigating the dynamics of single state vector we study the evolution equation  the density operator $\rho_{\alpha\alpha}=\ket{\nu_{\alpha}}$, we use $\ket{\nu_{\alpha}}\bra{\nu_{\alpha}}$ for pure states and $\rho_{\alpha\beta}=\ket{\nu_{\alpha}}\bra{\nu_{\beta}}$ for the mixed state, that depicts an ensemble of bi-flavored neutrino system. The state $\ket{\nu_{\alpha}}$ is from Hilbert space $\mathcal{H}_{1}$ and the state $\ket{\nu_{\beta}}$ is from Hilbert space $\mathcal{H}_{2}$. The dynamics of the combined system $\rho=\ket{\nu_{\alpha}}\bra{\nu_{\beta}}$ is defined over the Hilbert space $\mathcal{H}_{1}\otimes\mathcal{H}_{2}$\cite{Blasone:2019chv} which has the dimension of $\mathcal{H}_{2}\otimes\mathcal{H}_{2}$. The manifestation of this idea makes our choice basis more tangible. The density operator  $\ket{\nu_e}\bra{\nu_e}$ for pure electron neutrino state is,
\begin{equation}\label{eq:14}
\begin{split}
     \ket{\nu_e}\bra{\nu_e}=\rho_{ee}^{m} & =\cos^{2}{\theta}\ket{\nu_1}\bra{\nu_1}+\sin{\theta}\cos{\theta}\ket{\nu_1}\bra{\nu_2}\\
     & +\sin{\theta}\cos{\theta}\ket{\nu_2}\bra{\nu_1}+\sin^{2}{\theta}\ket{\nu_2}\bra{\nu_2}\\
     & =\cos^{2}{\theta}\ket{00}+\cos{\theta}\sin{\theta}\ket{01}\\
     & +\cos{\theta}\sin{\theta}\ket{10}+\sin^{2}{\theta}\ket{11}
\end{split}
\end{equation}
 While propagation, the flavor state is mixed with the mass eigenstate, and since each flavor state consists of two mass eigenstates, the purity of the system is not conserved, and the neutrino oscillates from one flavor state to the other.  Redefined mass eigen state consists of $\ket{00}=\ket{0}_{s}\otimes\ket{0}_{e}$ , $\ket{11}=\ket{1}_{s}\otimes\ket{1}_{e}$ ,$\ket{11}=\ket{0}_{s}\otimes\ket{1}_{e}$ and $\ket{10}=\ket{1}_{s}\otimes\ket{0}_{e}$. From eq.(\ref{eq:14}), the density operator denoting the flavor state,
 $\rho_{ee}=\begin{pmatrix}
        \cos^{2}{\theta} & \cos{\theta}\sin{\theta}\\
        \cos{\theta}\sin{\theta} & \sin^{2}{\theta}\\
    \end{pmatrix}$. We exploit the fact that $\mathcal{R}^{2}\otimes\mathcal{R}^{2}$ is isomorphic to $\mathcal{R}^{4}$ to vectorise the density operator. 
    The objective is to vectorize the Hamiltonian and density operator over a common basis to study the dynamics pictorially. This will give us a comparative understanding of the neutrino evolution and the precision of an NMR in an external magnetic field. To express the vectorized form of the density operator, we introduce a momentum-like vector $\Vec{F}$ such that $F_{i}=\mbox{Tr}\big[\frac{1}{2} \sigma_{i}\rho\big]$, where $i$ stands for the 1, 2, and 3 components.
     %The density vector is then expressed as $\begin{pmatrix}
     %   \cos^{2}{\theta}\\
      %  \cos{\theta}\sin{\theta}\\
       % \cos{\theta}\sin{\theta}\\
        %\sin^{2}{\theta}\\
    %\end{pmatrix}$ 
\begin{equation}\label{3.5}
\begin{split}
     F_{1} & =\mbox{Tr}\bigg[\frac{1}{2}\sigma_{1}\rho\bigg]\\
    & =\mbox{Tr}\bigg[\frac{1}{2}\begin{pmatrix}
        0 & 1\\
        1 & 0\\
    \end{pmatrix}\begin{pmatrix}
        \rho_{11} & \rho_{12}\\
        \rho_{21} & \rho_{22}\\
    \end{pmatrix}\bigg]\\
    & =\frac{1}{2}(\rho_{12}+\rho_{21})
\end{split}
\end{equation}
Similarly $F_{2}=\frac{i}{2}(\rho_{12}-\rho_{21})$ and $F_{3}=\frac{1}{2}(\rho_{11}-\rho_{22})$. Hence, the density matrix is constructed as,
\begin{equation}\label{eq:16}
    \rho_{ij}=\begin{pmatrix}
        F_{3} & F_{1}-iF_{2}\\
        F_{1}+iF_{2} & -F_{3}\\
    \end{pmatrix}
\end{equation}
$\rho_{ij}$ stands for the components of the density operator in Pauli bases. Hence, $\mathbf{F}$ signifies a vector that stands for the flavor state of the system that aligns toward the flavor basis. We consider the neutrinos created a pure electron flavor state. The two flavor states corresponding to the bi-flavored neutrino oscillation are orthogonal to each other. Accordingly, we take the $\nu_{e}$ state along $\sigma_{3}^{F}$ and $\nu_{\mu}$ along $-\sigma_{3}^{F}$. In flavor basis at $t=0$, $F_{1}^{F}=0$, $F_{2}^{F}=0$ and $F_{3}^{F}=1$. We move $\mathbf{F}$ from flavor to mass basis to study the evolution equation to avoid mathematical ambiguity. However, time evolved $F^{F}(t)$ will be applicable for understanding the scenario at the probability level.
Constructing so, we move to evaluate the evolution of the density operator. 
\\Now we have two vectors $\mathbf{F}$ and $\mathbf{B}$ representing the two operators, density operator $\rho$ and the total Hamiltonian including decay $\mathcal{H}_{m}$ respectively, defined over the Pauli basis.
The Hamiltonian vector $\mathbf{B}$ has a Hermitian component, $\mathbf{B}_{h}$, which is associated with mass square differences and a non-hermitian component, $\mathbf{B}_{nh}$, which incorporates the decay component of $\mathbf{B}$. 
Although we initiated the study of the Schrodinger equation, which states the dynamics of the flavor state, as soon as we introduced the concept of density operator, we moved from the Schrodinger picture to Heisenberg's picture, where the operator has the dynamical evolution instead of a state vector. 
%A point to be mentioned here is that whenever we have introduced the concept of density operator, we move to Heisenberg's picture from the Schrodinger picture
\begin{equation}\label{eq:17}
    i\frac{d}{dt}\rho\big(t\big)=\bigg(\mathbf{B}_{h}\rho(t)-\rho(t)\mathbf{B}_{h}\bigg)+\bigg(\mathbf{B}_{nh}\rho(t)-\rho(t)\mathbf{B}_{nh}^{\dagger}\bigg)
\end{equation}
For the non-hermitian nature of the decay Hamiltonian $\mathbf{B}_{nh}\neq\mathbf{B}_{nh}^{\dagger}$. However, discrepancies arise as the Hamiltonian is vectored but not the density operator. To rectify this inconsistency, we convert the density operator $\rho$ to the vectorized form mentioned previously by the vector $\Vec{F}$. Using the form of $\rho$ from eq.(\ref{eq:16}) the form of LHS of eq.(\ref{eq:17}) modified to,
\begin{equation}\label{eq:18}
\begin{split}
    \frac{d}{dt}\rho & =\begin{pmatrix}
        0 & -2i\omega(F_{1}-iF_{2})\\
        2i\omega(F_{1}+iF_{2}) & 0\\
    \end{pmatrix}\\
    & -2b\begin{pmatrix}
        F_{3} & F_{1}-iF_{2}\\
        F_{1}+iF_{2} & -F_{3}\\
    \end{pmatrix}\\
    & +\begin{pmatrix}
        -\eta F_{1} & 0\\
        0 & -\eta F_{1}\\
    \end{pmatrix}\\
\end{split}
\end{equation}
The RHS of the above equation changes its form following the form $\frac{d}{dt}F_{i}=Tr\frac{1}{2}\sigma_{i}\frac{d}{dt}\rho$. The component wise evolution equation of $\mathbf{F}$ is thus,
\begin{equation}\label{eq:19}
\begin{split}
     \Dot{F}_{1} & =-\omega F_{2}-b F_{1}\\
    \Dot{F}_{y} & =\omega F_{1}-b F_{2}\\
    \Dot{F}_{2} & =-2b F_{3}
\end{split} 
\end{equation}
In matrix form,
\begin{equation}\label{eq:20}
\begin{pmatrix}
    \Dot{F}_{1}\\
    \Dot{F}_{2}\\
    \Dot{F}_{3}\\
\end{pmatrix} =\begin{pmatrix}
    -b & -\omega & 0\\
    \omega & -b & 0\\
    0 & 0 & -2b\\
\end{pmatrix}\begin{pmatrix}
    F_{1}\\
    F_{2}\\
    F_{3}\\
\end{pmatrix}
\end{equation}
it is noteworthy that although we have vectorized the density operator as $\mathbf{F}$, we do not mention explicitly on which basis it is defined. In general, $\mathbf{F}$ illustrates the density operator of the flavor state of neutrino. We can denote flavor part of the  $\mathbf{F}$ as $\mathbf{F}_{i}^{F}$ while in mass part $\mathbf{F}_{i}^{M}$. As we consider neutrino is created as the pure electron flavor, hence at $t=0$, $\mathbf{F}_{1}^{F}=0$,$\mathbf{F}_{2}^{F}=0$, $\mathbf{F}_{3}^{F}=1$. In mass basis at $t=0$, $\mathbf{F}_{1}^{M}=\sin{2\theta}$,$\mathbf{F}_{2}^{M}=0$, $\mathbf{F}_{3}^{M}=\cos{2\theta}$. On that account eq.(\ref{eq:20}) donates the evolution equation of $\mathbf{F}^{M}$ in order to find time evolved $\mathbf{F}^{M}$ at $t=0$ \cite{Giunti:2007ry}. Rewriting eq.(\ref{eq:20}),
\begin{equation}\label{eq:21}
\begin{pmatrix}
    \Dot{F}_{1}^{M}\\
    \Dot{F}_{2}^{M}\\
    \Dot{F}_{3}^{M}\\
\end{pmatrix} =\begin{pmatrix}
    -b & -\omega & 0\\
    \omega & -b & 0\\
    0 & 0 & -2b\\
\end{pmatrix}\begin{pmatrix}
    F_{1}^{M}\\
    F_{2}^{M}\\
    F_{3}^{M}\\
\end{pmatrix}
\end{equation}
Recall we have mentioned earlier that $\omega=\frac{\Delta m^{2}}{4E}$. Accordingly, $2\omega=\frac{\Delta m^{2}}{2E}$, which defines the standard oscillation frequency of the flavor oscillation of neutrinos. Thus, we redraft $2\omega$ as $\omega_{osc}$ for mathematical simplification. It is to be noted that the time evolution of the density operator is not independent of $\eta$, i.e., the off-diagonal part of the decay Hamiltonian, which is also referred to in eq.(\ref{eq:18}). However, the time evolution of $\mathbf{F}$, i.e., vector depicting the flavor state, is independent of the off-diagonal element $\eta$. Choosing the Pauli basis makes the evolution equation independent of the off-diagonal element.

%%%%%%%%%%%%%%%%%%%%%%%%%%%%%%%%%%%%%%%%%%%%%%%%%%%%%%%%%%%%%%%%%%%%%
%%%%%%%%%%%%%%%%%%%%%%%%%%%%%%%%%%%%%%%%%%%%%%%%%%%%%%%%%%%%%%%%%%%%%
\section{Results and Discussions}
In this section, we discuss the evolution of $\mathbf{F}^{M}$ analytically and illustrate the result geometrically. We interpret the dynamical equation as a second-order differential equation, an easier way to understand the oscillation pictorially. We scrutinize two different scenarios. Firstly, we consider the damping parameters zero and give the geometrical representation figuratively. Next, we include the decay parameters to reduce the differential equation of neutrino oscillation to the modified form and represent it geometrically.

\begin{itemize}
    \item Scenario 1: We first consider the standard neutrino oscillation without including decay parameters. For the standard neutrino oscillation, we set $\eta=0,b=0$. Hence, from the eq.(\ref{eq:21}), we can observe that the evolution equation of the vector describing the density matrix $\mathbf{F}$ becomes $\Dot{F}_{1}^{M}=-\omega_{osc} F^{M}_{2}$, $\Dot{F}_{2}^{M}=\omega_{osc} F^{M}_{1}$ and $\Dot{F}_{3}^{M}=0$. Individually $F_{1}^{M}$ and $F_{2}^{M}$ satisfy the equation of simple harmonic oscillator. Hence, in compact form
   \begin{equation}\label{eq:22}
       \Ddot{F}_{1,2}^{M}=-\omega_{osc}^{2}F_{1,2}^{M}
   \end{equation}
However, to have a more rigorous mathematical picture, it will be convenient to express them in terms of $\mathbf{B}$ and $\mathbf{F}$, although the in eq.(\ref{eq:22}) $\omega_{osc}$ is nothing but the coefficient of $\mathbf{B}$ itself. Eq.(\ref{eq:21}) can be expressed as, $\Dot{\mathbf{F}}_{1}^{M}  =-\omega_{osc} F_{2}^{M}\Hat{\sigma}_{1}$ and $\Dot{\mathbf{F}}_{2}^{M} =\omega_{osc} F_{1}^{M}\Hat{\sigma_{2}}$, which is contracted to the form,
\begin{equation}\label{eq:23}
\begin{split}
    \Dot{\mathbf{F}} & =-\omega_{osc} F_{2}^{M}\Hat{\sigma_{1}}+\omega_{osc} F_{1}^{M}\Hat{\sigma}_{2}\\
    & =\bigg(\omega_{osc}\Hat{\sigma}_{3}\times F^{M}_{2}\Hat{\sigma}_{2}\bigg)+\bigg(\omega_{osc}\Hat{\sigma}_{3}\times F_{1}\Hat{\sigma}_{1}\bigg)\\
    & =\mathbf{B}\times\mathbf{F}\\
\end{split}
\end{equation}
%\begin{equation}\label{eq:22}
 %    \frac{d}{dt}\mathbf{F}  =\mathbf{B}\times\mathbf{F}
%\end{equation}
  \begin{figure*}[htb!]
\centering
\includegraphics[scale=0.40]{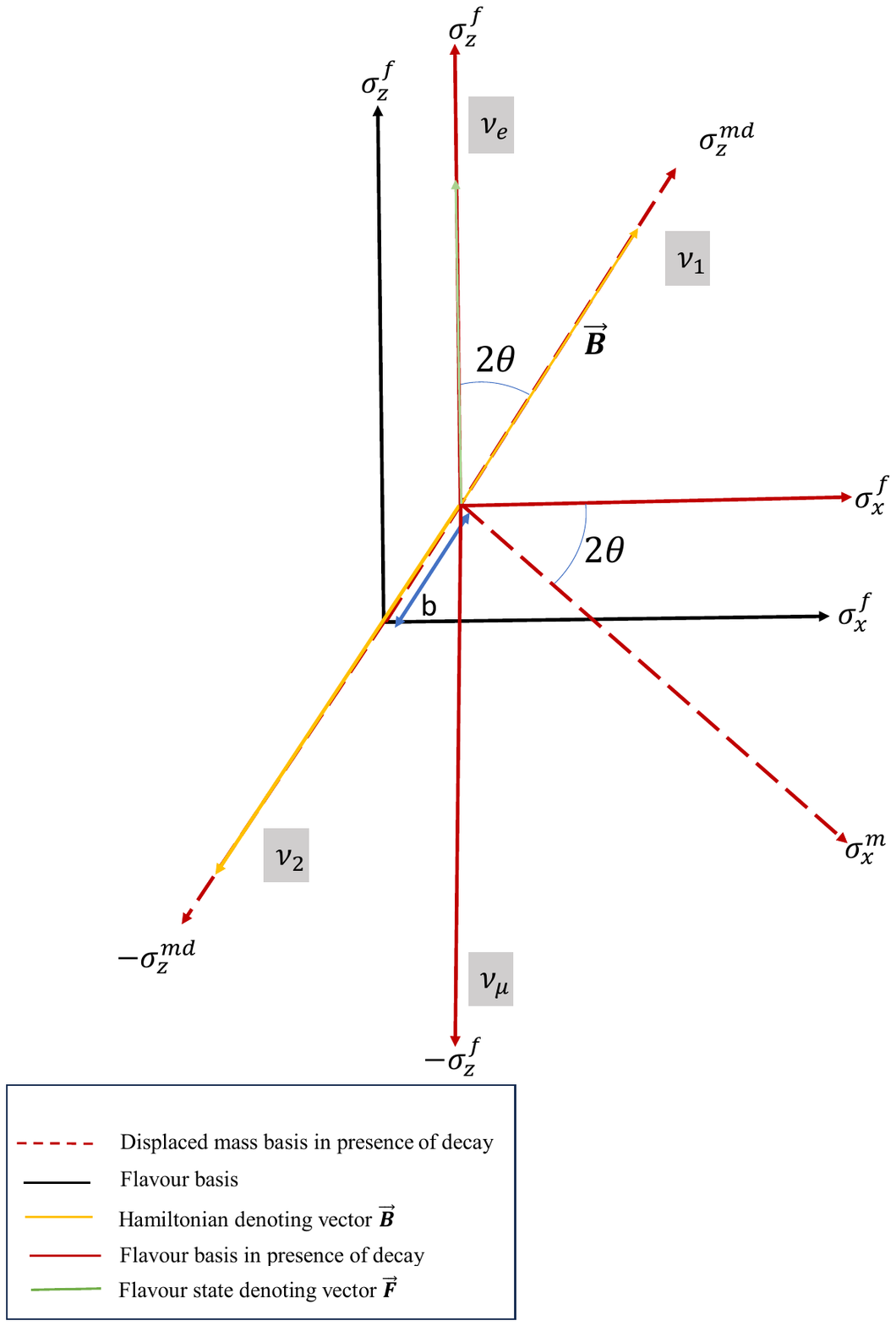}
\includegraphics[scale=0.40]{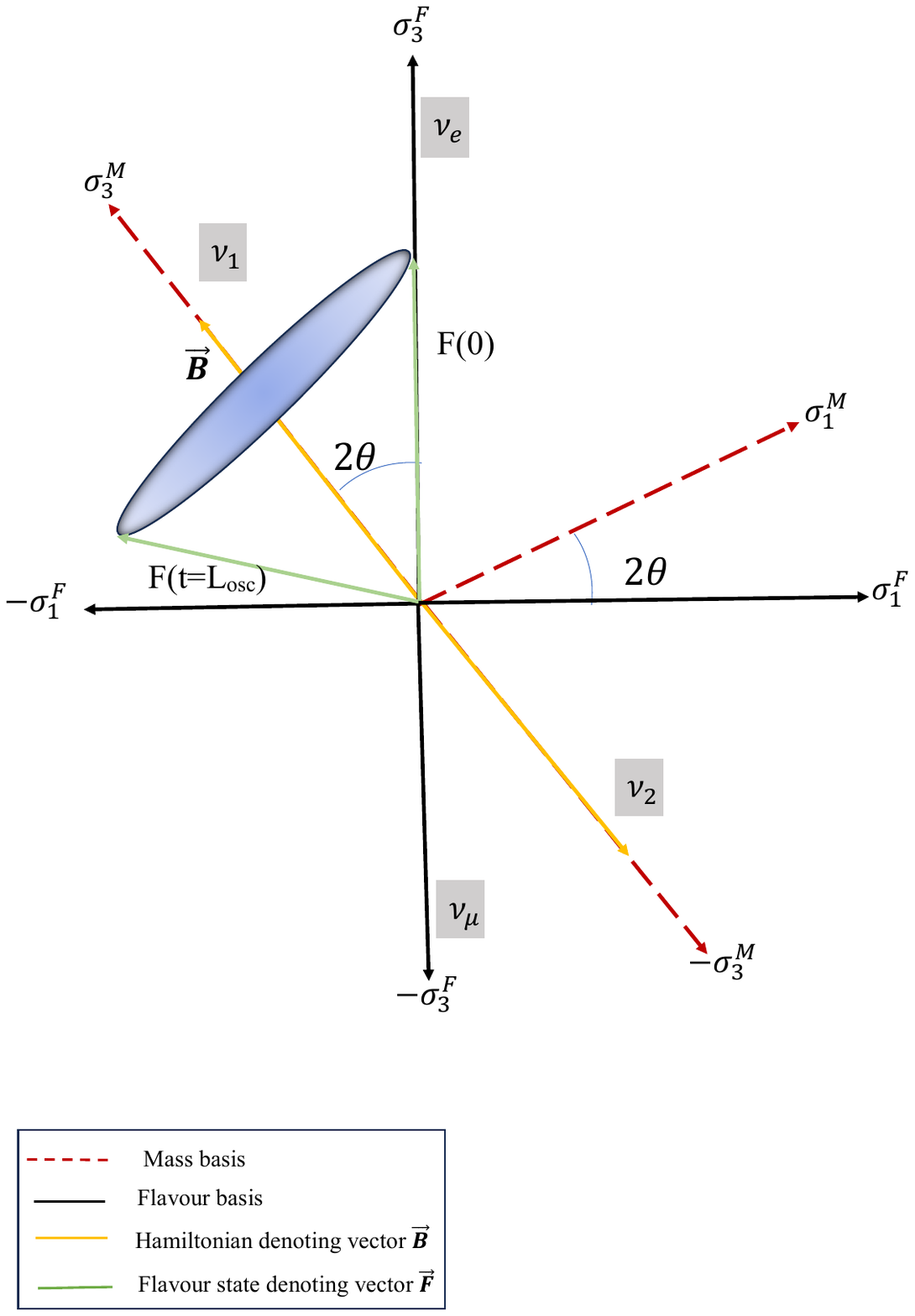}
\vspace*{0 cm}
\caption{Geometrical representation of basis in flavor basis and mass basis in the presence of decay(\textbf{left-panel}) and geometrical representation of standard neutrino oscillation without decay(\textbf{right-panel}). Here we follow the Pauli basis, i.e., $\Hat{\sigma}_{1}$, $\Hat{\sigma}_{2}$, $\Hat{\sigma}_{3}$ as the basis vector by which we define the coordinate axes. Plot legends in the box define the color representation of different components of the illustration. 
As the diagonal terms of the decay Hamiltonian match with the mass bases, it acts as a displacement element to the $\Hat{\sigma}_{3}$ component of the mass basis. The Hamiltonian depicting vector $\Vec{\mathbf{B}}$ is along $\Hat{\sigma}_{3}$. The $\sigma_{3}$ direction denotes the $\nu_{1}$ mass eigenstate and $-\sigma_{3}$ directs toward the $\nu_{2}$. On the other hand, in the flavor basis, the two orthogonal flavor states $\nu_{e}$ and $\nu_{\mu}$ align along positive and negative $\sigma_{3}$ directions, respectively, as illustrated in the left figure. The figure in the right panel shows that in the absence of decay, the flavor state representing vector $\Vec{\mathbf{F}}$ rotates around the Hamiltonian vector $\Vec{\mathbf{B}}$ following in the equation $\Dot{\mathbf{F}}=\mathbf{B}\times\mathbf{F}$. The initial flavor state is of electron neutrino; hence, $\mathbf{F}$ is along $+\Hat{\sigma}_{3}$. After a finite time $t$, $\mathbf{F}$ is along $\mathbf{F}(t)$, which implies the mixed state. After one complete revolution, it comes back to the initial pure state. In the relativistic limit $t\equiv L$, the angular velocity of the rotation is equivalent to the oscillation length $L_{osc}$, shown in the adjacent figure on the right.}
\label{fig:1}
\end{figure*}
This leads to the standard form of the Bloch equation with oscillation frequency $\omega_{osc}=\Delta m^{2}/2E$. Also, eq.(\ref{eq:23}) implies the gyromagnetic ratio is 1 for this precessional motion of neutrinos. 
The diagram in the right panel of fig.\ref{fig:1} figuratively refers to this precessional motion. At time $t=0$, the initial electron flavor state representing vector $\mathbf{F}$(green line) is aligned along the $\sigma_{3}^{F}$ and is denoted by $F(0)$. After a certain time $t$, the $\mathbf{F}$ will rotate to a position and be aligned to the position $F(t)$. Considering the ultra-relativistic limit for the neutrinos $t$ can be substituted with the oscillation length. After the completion of one Oscillation length, $\mathbf{F}$ will acquire the state $F(0)$. In the absence of decay Hamiltonian, this circular motion of $\mathbf{F}$ around $\mathbf{B}$(yellow line) will remain unaffected. Flavor and the mass basis are mentioned in the plot legend.
Bloch equation represents the evolution equation of the spin-angular moment vector or magnetic moment vector along a direction of the externally applied magnetic field. Analogically, it can be inferred that the Hamiltonian in the mass basis acts as a magnetic field, and the flavor state revolves around it with the precessional frequency $\omega$. 

    % \begin{figure*}[htb!]
%\centering
%\includegraphics[scale=0.40]{decay 1.pdf}
%\includegraphics[scale=0.40]{decay 2.pdf}
%\vspace*{-3 cm}
%\caption{Plot of variation of eigenvalues with energy(GeV) in vacuum without parameterising the decay Hamiltonian}
%\label{plot:unifn_wo_threshold}
%\end{figure*}

The applied magnetic field acts as a perturbation to the system, which exerts a torque in the equilibrium condition of the system, driving the system to rotate around. A similar phenomenon occurs for the system of neutrino as well, where the Hamiltonian, which contains the squared difference term, illustrates the linear combination of the two different masses; one is the system mass, and the other is the mass acquired from the vacuum while propagation.
\begin{figure*}[htb!]
\centering
\includegraphics[scale=0.50]{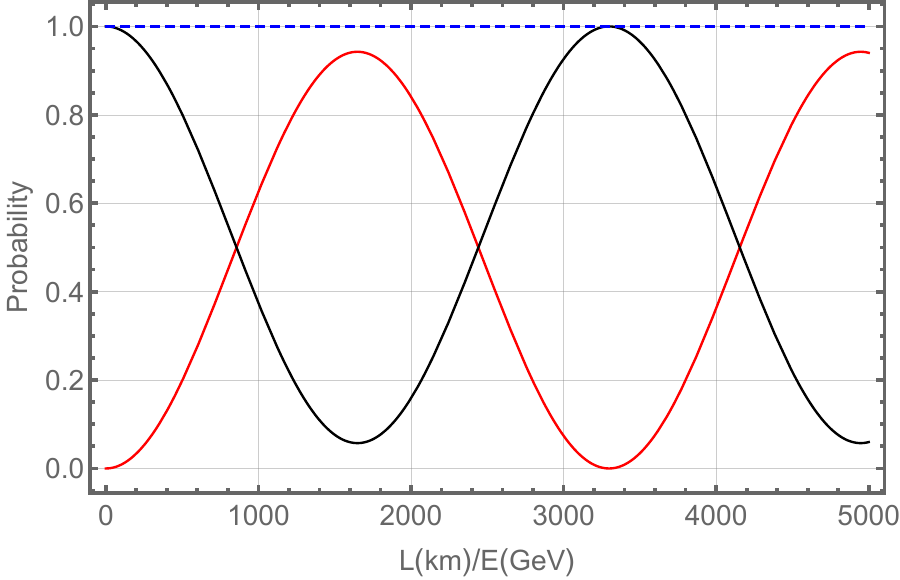}
\hspace{0.5cm}
\includegraphics[scale=0.45]{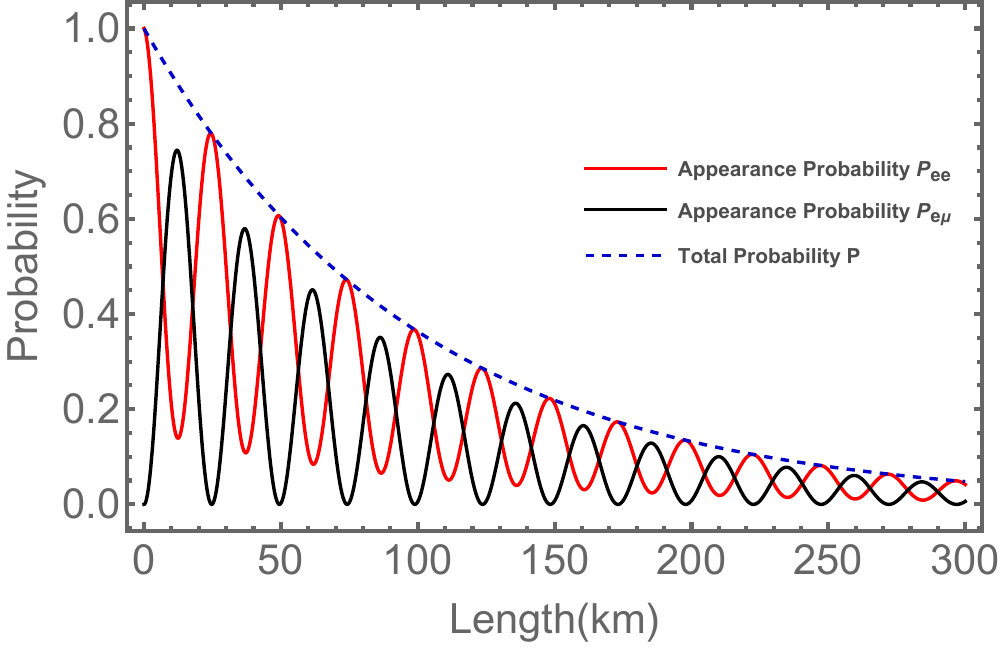}
\caption{Plot of appearance probability $P_{ee}$(red line), disappearance probability $P_{e\mu}$(black line) and the total probability $P$(Blue-dashed line). The plot in the right panel shows the probabilities with the variation of $L(km)/E(GeV)$ in vacuum. In the absence of decay, the two-flavor state acts as a two-level system with a constant oscillation amplitude, which keeps oscillating without deviation. Meanwhile, the figure in the right panel shows the probabilities in the presence of decay. Graphically it describes that the decay Hamiltonian which coincides with the mass basis is responsible for the amplitude damping of probabilities according to the expression $P_{ee}^{d}=e^{-2bL}P_{ee}$, $P_{e\mu}^{d}=e^{-2bL}P_{e\mu}$, $P^{d}=e^{-2bL}$. The plot legends mention the figure indicates the appearance, disappearance, and total probability variation. The figure manifests the fact that with the increasing length $L(km)$, the amplitude of a $P_{ee}^{d}$, $P_{e\mu}^{d}$ decreases in an oscillatory manner while the total probability $P^{d}$ decreases exponentially, with $E=1GeV$. This figure also conveys the graphical layout of under-damped oscillation when the decay parameter magnitude of the decay parameter $(b)$ is much less than angular velocity $\omega L$. The value of the parameter $b$ is $10^{-21}eV$ for considering $1eV=\Delta m^{2}$\cite{Dixit:2022izn} }
\label{fig:2}
\end{figure*}
%\begin{figure*}[htb!]\label{fig:4}
%\centering
%\includegraphics[scale=0.38]{modi_crit_dis.pdf}
%\includegraphics[scale=0.39]{modi_crit_app.pdf}
%\hspace{5cm}
%\includegraphics[scale=0.38]{modi_over_dis.pdf}
%\includegraphics[scale=0.41]{modi_over_app.pdf}
%\caption{The Figure demonstrates a comparative study of the critically damped(two plots in the left panel) and over damped(two plots in the left panel) neutrino oscillation. The plot legends indicate the appearance and disappearance probabilities in both cases. In the critically damped the decay parameter $b\equiv \frac{1.27\omega L}{2}$. In this case, the peak of disappearance probability is much less, although it is not completely negligible. In the scenario of over-damped case $b>\frac{1.27\omega L}{2}$, which shows a completely negligible disappearance probability with an exponential decay of the appearance probability. In this case, they adhere to the absolute decay of the pure state. }
%\label{plot:unifn_wo_threshold}
%\end{figure*}

Usually, this phenomenon cannot be seen when a particle has a definite mass. A single flavor state consisting of a single mass eigenstate may not possibly execute the oscillation. This adds the unique dynamical feature of eq.(\ref{eq:23}), making it distinct from any other fermions. $\ket{\nu_{\alpha}}=c_{1}\ket{\nu_{1}}+c_{2}\ket{\nu_{2}}$. This torque appears due to the superposition of two mass eigenstates. In addition to this, the existence of different flavor states makes the phenomenon more prominent. Illustratively, it is described by Fig.\ref{fig:1}. We 
 have considered that neutrino is created as a pure electron flavor state. In flavor basis, the pure electron flavor state is lined up along $\sigma^{f}_{3}$, and the muon flavor state is along $-\sigma^{f}_{3}$. Considering the transformation of the basis vectors, the mass and the flavor basis are inclined by an angle $2\theta$. This is known as the mixing angle of neutrino. eq.(\ref{eq:13}) suggests that in mass basis $\mathbf{B}$ is aligned along $\sigma^{M}_{3}$ with the states $\nu_{1}$ in $\sigma^{M}_{3}$ and $\nu_{1}$ in $-\sigma^{M}_{3}$. Imposing the initial condition at $t=0$, i.e., in flavor basis $\mathbf{F}_{1}^{F}=0$,$\mathbf{F}_{2}^{F}=0$, $\mathbf{F}_{3}^{F}=1$ and in mass basis, $\mathbf{F}_{1}^{M}=\sin{2\theta}$,$\mathbf{F}_{2}^{M}=0$, $\mathbf{F}_{3}^{M}=\cos{2\theta}$, the solution of eq.(\ref{eq:23}) is reduced to,
% or, more specifically, introducing more than one mass in a particular flavor state perturbates the system. Eq.(\ref{eq:22}) also matches with \cite{Giunti:2007ry}.
% Following the solution of \cite{Giunti:2007ry}, the density vector in the mass basis becomes,
 \begin{equation}\label{eq:24}
\begin{split}
    F_{1}^{M}(t) & =F_{1}^{M}(0)\cos{\omega_{osc} t}+F_{2}^{M}(0)\sin{\omega_{osc} t}\\
    F_{2}^{M}(t) & =F_{1}^{M}(0)\cos{\omega_{osc} t}-F_{2}^{M}(0)\sin{\omega_{osc} t}\\
    F_{3}^{M}(t) & =F_{3}^{M}(0)\\
\end{split}
 \end{equation}
 Which follows, $F^{M}_{1}=-\sin{2\theta}\cos{\omega_{osc} t}$, $F^{M}_{2}=-\sin{2\theta}\sin{\omega_{osc} t}$, $F^{M}_{3}=\cos{2\theta}$.
 So far, we are in the mass basis where the vector defining the density matrix is also defined over the mass basis. This refers to the solution of the evolution equation of $\mathbf{F}$ around $\mathbf{B}$ in mass basis. To understand how the dynamics affect the probability level $\mathbf{F}^{M}(t)$ has to be transformed into the flavor basis $\mathbf{F}^{F}(t)$ and $\mathbf{F}^{F}_{3}(t)$ will give the probability of obtaining the initial flavor $\nu_{e}$ after a time t. 
% But to investigate how these dynamics modify the flavor basis, we have to transform the vector defining the density matrix in the flavor basis. Since all these vectors are defined over the Pauli matrix basis, converting the basis vectors from the mass to the flavor basis will give the transformation of the desired vector. 
 \begin{figure*}[htb!]
 \centering
 \hspace*{0cm}
\includegraphics[scale=0.40]{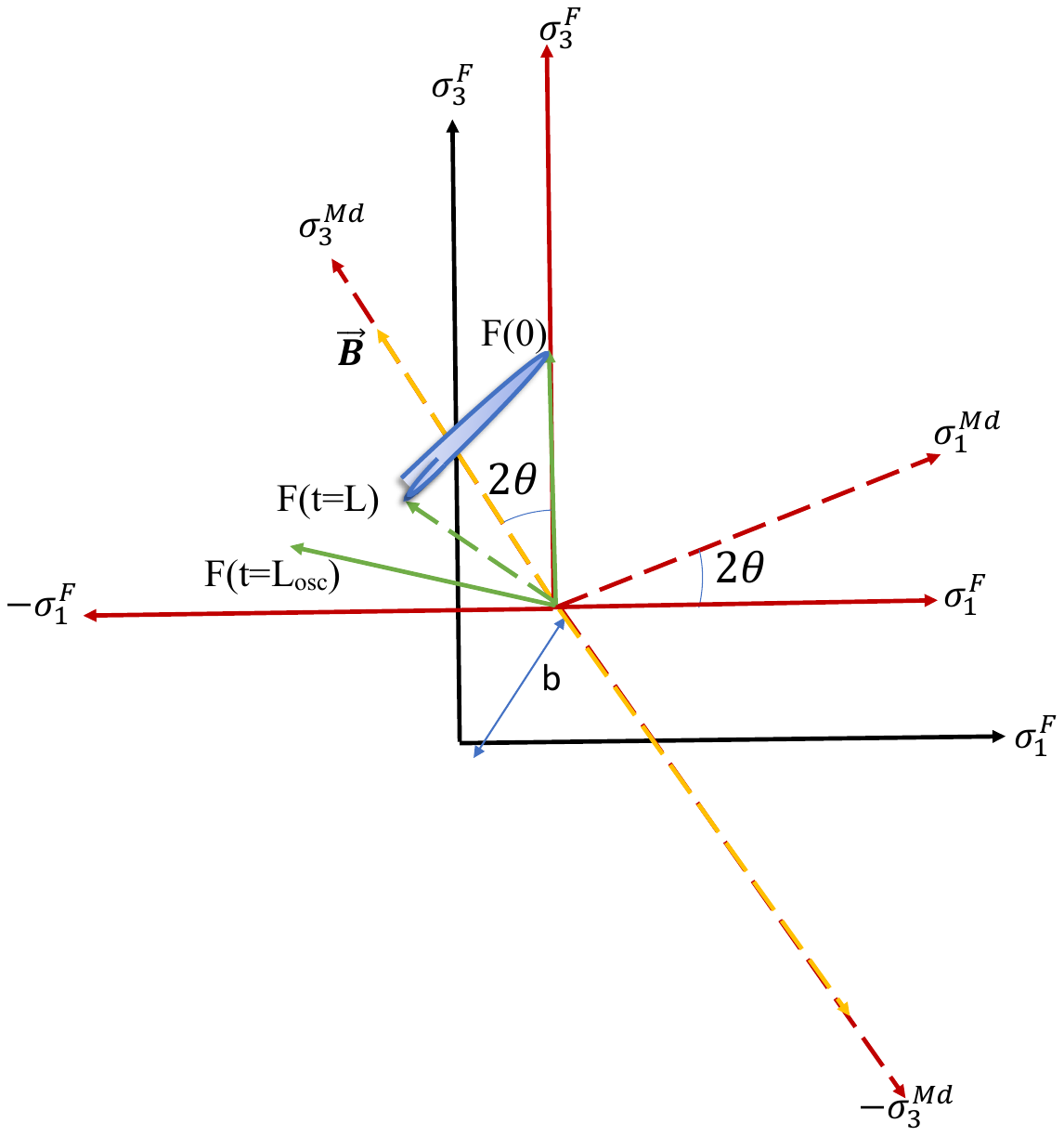}
\includegraphics[scale=0.40]{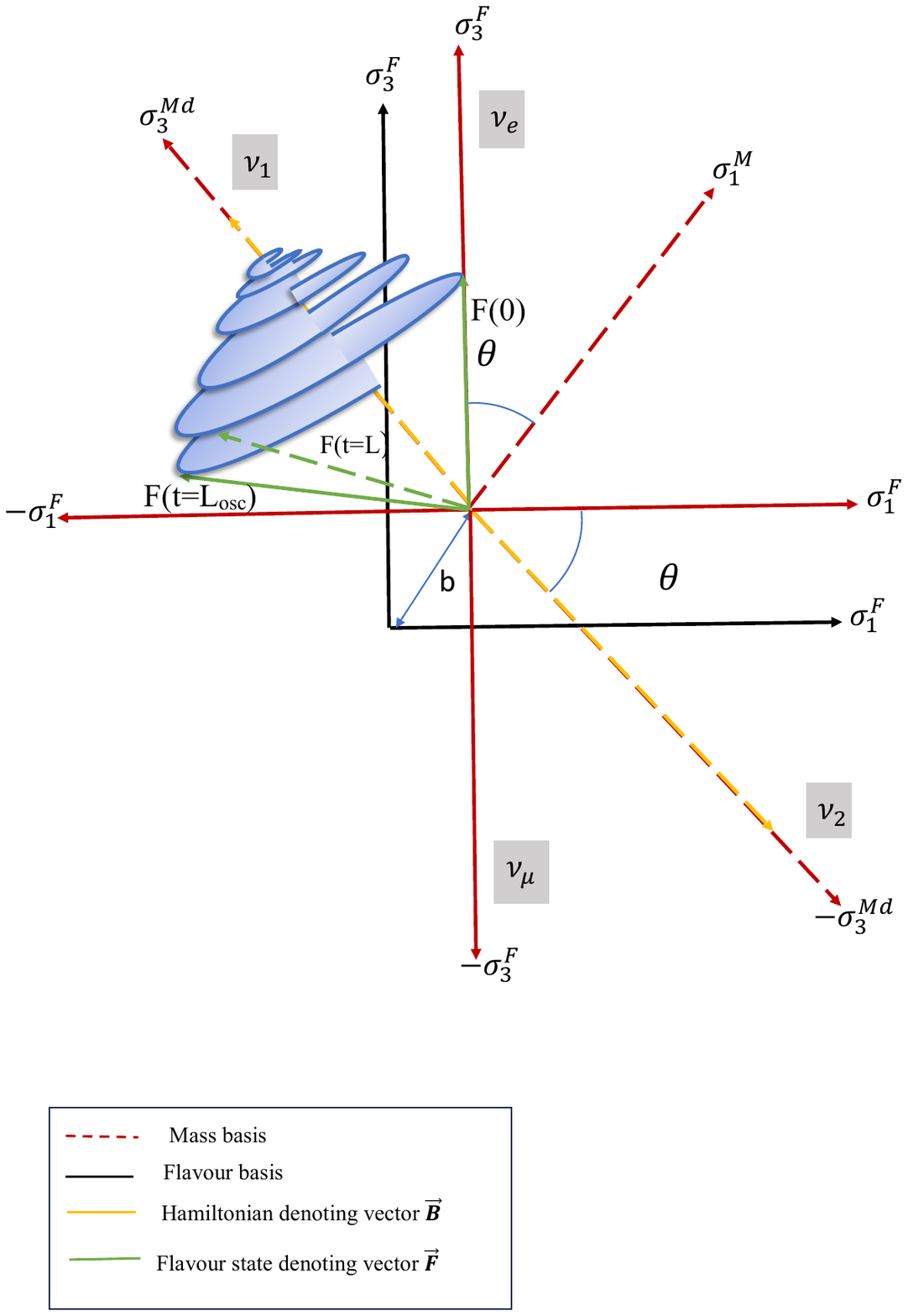}
\vspace*{0 cm}
\caption{Figures illustrate the geometrical representation of neutrino oscillation in the presence of decay for critically damped case(\textbf{left panel}) and under damped case(\textbf{right panel}). For the under-damped case, the oscillation amplitude damped down in an oscillatory way. Consequently, there is oscillation with decreasing amplitude, which directs towards the fact that after one full revolution, it is impossible to get back to the initial flavor state as the presence of decay parameter $b$, which causes the flavor state depicting vector $\mathbf{F}$ execute a helical motion. Moreover, this helical motion doesn't confine to a plane because this decay parameter affects the $\Hat{\sigma}_{3}$ components as well, which makes $\mathbf{F}$ drag toward $\Hat{\sigma}_{3}^{md}$ which the mass basis in the presence of decay. In addition, with the gradual decrement of the pure state, the flavor state dies down, demonstrating that it ultimately dissipates to the mass state and $\mathbf{F}$ aligns itself towards $\mathbf{B}$. In a critically damped case, the magnitude of the damping parameter is equal to the natural frequency of the oscillation. Due to this equivalence, the $\mathbf{F}$ vector is set off for rotational motion, although it can not complete one full rotation and decays to the mass eigenstate.}
%\label{plot:unifn_wo_threshold}
\label{fig:3}
\end{figure*}
 \begin{figure}[htb!]
 \centering
 \hspace*{0cm}
\includegraphics[scale=0.40]{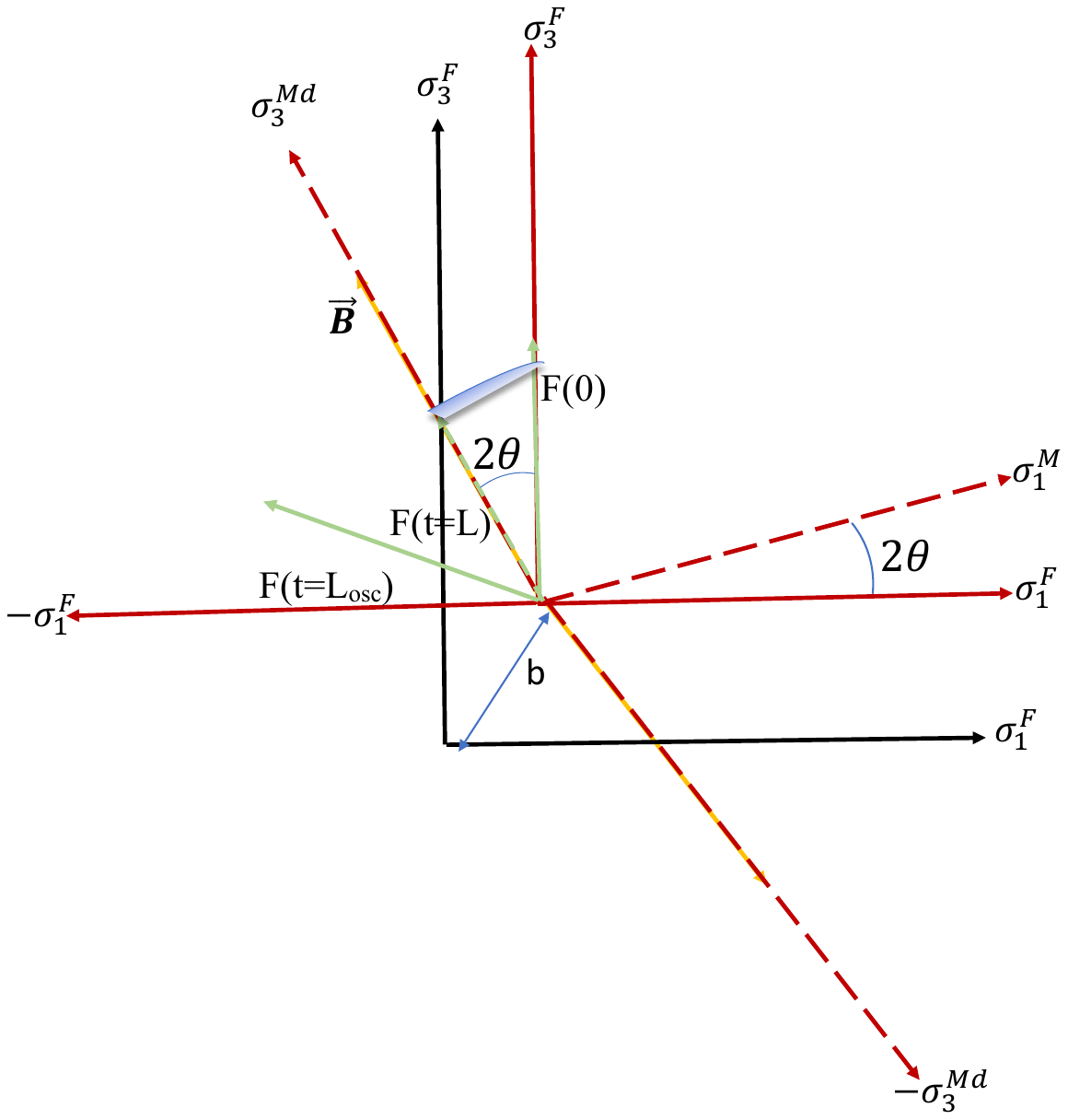}
\vspace*{-4 cm}
\caption{Figures illustrate the geometrical representation of neutrino oscillation in the presence of decay for the over-damped scenario. The oscillation amplitude damped down in an oscillatory way in the presence of decay. Consequently, there is oscillation with decreasing amplitude, which directs towards the fact that after one full revolution, it is impossible to get back to the initial flavor state as the presence of decay parameter $b$, which causes the flavor state depicting vector $\mathbf{F}$ execute a helical motion. Moreover, this helical motion doesn't confine to a plane because this decay parameter affects the $\Hat{\sigma}_{3}$ components as well, which makes $\mathbf{F}$ drag toward $\Hat{\sigma}_{3}^{md}$ which the mass basis in the presence of decay. In addition, with the gradual decrement of the pure state, the flavor state dies down, demonstrating that it ultimately dissipates to the mass state and $\mathbf{F}$ aligns itself towards $\mathbf{B}$. In a critically damped case, the magnitude of the damping parameter is equal to the natural frequency of the oscillation. Due to this equivalence, the $\mathbf{F}$ vector is set off for rotational motion, although it can not complete one full rotation and decays to the mass eigenstate. The scenario of decay of flavor state to the mass eigenstate is much more dominant for the over-damped case where $\mathbf{F}$ directly dissipates to $\mathbf{B}$ without executing any rotational motion as shown by the figure in the left panel.}
\label{fig:4}
\end{figure}
 Following the transformation, $\Hat{\sigma}^{F}_{3}=\cos{2\theta}\Hat{\sigma}_{3}^{M}-\sin{2\theta}\Hat{\sigma}_{1}^{M}$, of the basis vectors, the $\Vec{F}$ can be transformed into the flavor basis, i.e., $F_{i}^{F}(t)$. 
 %Considering initially the neutrino is in a pure state, i.e., generated in the pure flavor state, we represent the flavor state along the z-axis of the flavor basis. $+z$ axis denote the pure electron flavour and $-z$ axis denote the pure muon flavour. These two bases are related by an opening angle of $2\theta$ amongst each other. The density vector in flavour basis established by coordinate transformation $F_{z}^{f}=-\sin{2\theta}F_{x}^{m}+\cos{2\theta}F_{y}^{m}$. 
 We are only interested in the $\sigma_{3}$ component of the $\Vec{F}$ because the initial flavor state is along the $\Hat{\sigma_{3}}$ component. The time-evolved component of $F_{z}$ provides the flavor state, which eventually gives the appearance probability of the particular flavor state, i.e., $P_{\alpha}=F_{3}^{M}(t)=1-\sin^{2}{2\theta}\sin^{2}{\frac{\Delta m^{2}L}{4E}}$. The plot at the left of Fig.\ref{fig:2} depicts the same behavior of standard oscillation. The appearance (black line) and disappearance (red line) probabilities are plotted with respect to the $L(km)/E(GeV)$ ratio. This figure points to the un-deviated oscillatory nature of the precessional motion of neutrinos, mentioned in Fig.$\ref{fig:1}$ which also implies the same picturization of any standard two-level system \cite{Manzano:2020abc}.
 %Observing \ref{fig:1} and \ref{eq:21}, the flavor basis is rotating with respect to the Hamiltonian with angular frequency $\omega$, which depends on the oscillation length. 
 %\ref{eq:14} will be in the flavour basis as, $c_{0}^{f}$, $c_{x}^{f}$, $c_{y}^{f}$, $c_{z}^{f}$.
   %  Before showing the explicit form of the, we mention that initially, we have taken the neutrino produced in the pure electron state $\ket{\nu_{e}}$ which follows $c^{f}_{x}=c^{f}_{y}=0$ and $c^{f}_{z}=1$. Hence following the same transformation the time evolved flavor basis will be $c^{f}_{z}=\sin{2\theta}c^{m}_{1}+\cos{2\theta}c^{m}_{2}=1-2\sin^{2}{2\theta}\sin^{2}{\omega t}$. From \ref{eq:22}, the probability will be $P_{ee}=\rho_{ee}^{f}=\frac{1}{2}(1+c^{f})=1-\sin^{2}{2\theta}\sin^{2}{\omega t}$ which is the exact the same expression as the Appearance probability of the neutrino oscillation. The critical point is that we have reduced probability expression obtained just by transforming the coordinate in Liouville space from one basis to another.
   %  \begin{widetext}
% \begin{figure}[htb!]\label{fig:1}
%\centering
%\includegraphics[scale=0.50]{Prob_std.pdf}
%\hspace{0.1cm}
%\includegraphics[scale=0.50]{Prob_decay.pdf}
%\caption{Plot of a). b) .}
%\label{plot:unifn_wo_threshold}
%\end{figure}
%\end{widetext}
   It also predicts after which length we should get back the same flavor. Intuitively, this refers to the fact that the flavor state of the neutrino acts as a magnetic moment vector rotating along the magnetic field where the Hamiltonian depicts the magnetic field.
% \begin{figure*}[htb!]
%\centering
%\includegraphics[scale=0.20]{decay 1.pdf}
%\hspace{0.1cm}
%\includegraphics[scale=0.20]{decay 2.pdf}
%\hspace{0.1cm}
%\includegraphics[scale=0.55]{std_osc.pdf}
%\caption{Plot of variation of eigenvalues with energy(GeV) in vacuum without parameterising the decay Hamiltonian}
%\label{plot:unifn_wo_threshold}
%\end{figure*}
   
    \item Scenario 2: $\eta\neq0, b\neq 0$. In the next scenario, we analyze considering both the decay parameters to be non-zero of the decay Hamiltonian, which effectively modifies the evolution equation eq.(\ref{eq:19}) to,
   \begin{equation}\label{eq:25}
    \begin{split}
        \Ddot{F}_{1}^{M} & =\omega_{osc}^{2}F_{1}^{M}-2b\Dot{F_{1}}^{M}\\
        \Ddot{F}_{2}^{M} & =\omega_{osc}^{2}F_{2}^{M}-2b\Dot{F_{2}}^{M}\\
        \Dot{F}_{3}^{3} & =-2bF_{3}^{M}
    \end{split}
   \end{equation} 
   Analogically, the above equation is the equation of motion of a damped harmonic oscillator. Physically, one such instance can be noticed in NMR. In the case of NMR, the $2b$ factor acts as the relaxation time $1/T$. 
   It is noted from eq.(\ref{eq:25}) that the $\eta$ factor doesn't play a role here. If the off-diagonal term is hermitian, it contributes to the evolution equation, which can be noted following \cite{Am-Shallem_2015}.
   %Hence, to get the contribution of the decay factor, which doesn't coincide with the mass basis, the $\xi$ factor has to be taken into account. 
   For the invisible neutrino decay, the Bloch equation is independent of the diagonal terms of the perturbation Hamiltonian. Moreover, the contribution from the $\sigma_{3}$ component makes the equation different from the standard oscillation, where no contribution comes from the $\sigma_{3}$ component. To visualize the dynamics geometrically, let us look through the differential Bloch equation form of eq.(\ref{eq:24}).
   \begin{equation}\label{eq:26}
      \frac{d}{dt}\mathbf{F}  =\mathbf{B}\times\mathbf{F}-2b\mathbf{F}
   \end{equation}
   
 Eq.(\ref{eq:25}) implies that due to the presence of the decay factor, the circular motion is converted to a spiral motion coiling into the mass basis of the Hamiltonian. Moreover, due to the occurrence of damping along the $\sigma_{3}^{M}$-axis $\mathbf{B}$ drag $\mathbf{F}$, which enforced $\mathbf{F}$ to execute a helical motion in the $\sigma_{3}$ direction with gradually decreasing screw pitch. We must quantify parameter b with an approximated value mentioned in \cite{Dixit:2022izn, Soni:2023njf} for a complete pictorial overview. For consistency, $b$ has to be the dimension of energy(eV). Next, we derive the oscillation probability using eq.(\ref{eq:25}), the solution of which,
 \vspace{-0.1cm}
  \begin{equation}
\begin{split}
    F_{1}^{M}(t) & =e^{-2bt}F_{1}^{M}(0)\cos{\omega_{osc} t}\\
    & +e^{-2bt}F_{2}^{M}(0)\sin{\omega_{osc} t}\\
    F_{2}^{M}(t) & =e^{-2bt}F_{1}^{M}(0)\cos{\omega_{osc} t}\\
    & -e^{-2bt}F_{2}^{M}(0)\sin{\omega_{osc} t}\\
    F_{3}^{M}(t) & =e^{-2bt}F_{3}^{M}(0)\\
\end{split}
 \end{equation}
Here, we impose the same boundary condition as before,i.e., $F_{1}^{M}(0)=-\sin{2\theta}$, $F_{2}^{M}(0)=0$, $F_{2}^{M}(0)=\cos{2\theta}$. Transforming the basis vector mass to flavor basis following $\Hat{\sigma}^{F}_{3}=\cos{2\theta}\Hat{\sigma}_{3}^{M}-\sin{2\theta}\Hat{\sigma}_{1}^{M}$, furnish the expression of $F_{3}^{F}(t)$ which is the probability of detecting flavour state $\nu_{e}$. Explicit mathematics follows,
\begin{equation}
\hspace{-2cm}
    F_{3}^{F}(t)=\cos{2\theta}F^{M}_{3}-\sin{2\theta}F^{M}_{1}
\end{equation}
This produces the oscillation probability in the presence of decay Hamiltonian 
\begin{equation}
    F_{3}^{F}(t)=e^{-2bt}\bigg[1-\sin^{2}{2\theta}\sin^{2}\bigg({\frac{\omega_{osc}t}{2}}\bigg)\bigg]
\end{equation}
We have approximated $\omega_{osc}>>b$, so the oscillation frequency remains unchanged.
Nevertheless, the inclusion of $b$ modifies the oscillation to $\sqrt{\omega_{osc}^{2}+b^{2}}$. The expression refers to the phenomenon of amplitude damping of neutrino oscillation apart from imposing a perturbation to the mass basis. The plot at the right panel of Fig.$\ref{fig:2}$ implies the same fact, while the figure in the right panel of Fig.\ref{fig:3} implements the same factuality diagrammatically. 
\\A parallel comparison can be drawn with the NMR system. The nuclear spin system needs $T_{1}$ to reach the thermodynamic equilibrium in the presence of the applied transverse external magnetic field. This results in exponential damping. Moreover, the relaxation time depends on the material of the medium and the direction of the applied field \cite{Kim2008PulsedN}. The Bloch matrix for the NMR system can be followed from \cite{Am-Shallem_2015}, which shows.
\begin{equation}\label{eq:45}
\begin{pmatrix}
    \Dot{S}_{x}\\
    \Dot{S}_{y}\\
    \Dot{S}_{z}\\
\end{pmatrix} =\begin{pmatrix}
    -\frac{1}{T_{1}} & \Delta & 0\\
    -\Delta & -\frac{1}{T_{1}} & \epsilon\\
    0 & -\epsilon & -\frac{1}{T_{2}}\\
\end{pmatrix}\begin{pmatrix}
    S_{x}\\
    S_{y}\\
    S_{z}\\
\end{pmatrix}+\begin{pmatrix}
    0\\
    0\\
    S_{z}^{0}\\
\end{pmatrix}
\hspace{-0.5cm}
\end{equation}
Dependence of the relaxation time over the direction of the applied magnetic field introduces the two relaxation times $T_{1}$ and $T_{2}$. Equivalently, the system of the decay parameter defines the lifetime of the flavor state \cite{Dixit:2022izn, Chattopadhyay:2021eba} that follows a straightforward analytic definition of the decay parameter, which is subjected to a dependence on the mass of the neutrino $m_{\nu}$ and lifetime of neutrino $\tau_{\nu}$. Since no definite magnetic field applies to the neutrino system, only one damping factor $b$ exists. 
\item Scenario 3: $\eta\neq 0, b\neq 0, \xi\neq 0$. This is the most general scenario where we do not approximate the $\xi$ to be zero for the invisible neutrino decay. Inclusion of $\xi$ term in the off-diagonal part of the decay Hamiltonian further bifurcates $\mathbf{B}_{nh}$ in the following way,
\begin{equation}
    \mathbf{B}_{nh}= \frac{-i}{2}(\sigma_{0}b)-\frac{-i}{4}\cos{\xi}\Hat{\sigma}_{1}+\frac{i}{4}\sin{\xi}\Hat{\sigma}_{2}
\end{equation}
In our present choice of basis, again, the $\Hat{\sigma}_{1}$ component will be decoupled from the dynamics of $\Dot{F}^{M}$. However, a contribution remains from the $\Hat{\sigma}_{2}$ component. From the commutation relation $\frac{d}{dt}\rho=-\frac{\sin{\xi}}{2i}\begin{pmatrix}
    -F_{1} & F_{3}\\
    F_{3} & F_{1}\\
\end{pmatrix}$ which modifies the the evolution equation of $\mathbf{F}^{M}$ to,
\begin{equation}\label{eq:32}
\begin{split}
    \Dot{F}_{1} & =-\omega F_{2}-bF_{1}+\eta\frac{\sin{\xi}}{2i}F_{3}\\\
    \Dot{F}_{2} & =\omega F_{1}-bF_{2}\\
    \Dot{F}_{3} & =-2bF_{3}-\eta\frac{\sin{\xi}}{2i}F_{1}
\end{split}
\end{equation}
In more compact form, the matrix representation of eq.(\ref{eq:32}) is written as,
\begin{equation}
    \begin{pmatrix}
    \Dot{F}_{1}^{M}\\
    \Dot{F}_{2}^{M}\\
    \Dot{F}_{3}^{M}\\
\end{pmatrix} =\begin{pmatrix}
    -b & -\omega & \eta\frac{\sin{\xi}}{2i}\\
    \omega & -b & 0\\
   -\eta\frac{\sin{\xi}}{2i}  & 0 & -2b\\
\end{pmatrix}\begin{pmatrix}
    F_{1}^{M}\\
    F_{2}^{M}\\
    F_{3}^{M}\\
\end{pmatrix}
\end{equation}
The above equation indicates that the inclusion of the $\xi$ term can not decouple the $\eta$ term from the derivation, which is a clear-cut indication of the presence of $CP$ phase even in the absence of the Majorana phase $\phi$. The effect of the factor $\xi$ in a more explicit way can be shown in the following way.
\begin{equation}\label{eq:34}
\begin{split}
    \Dot{\mathbf{F}} & =\bigg(\omega_{osc}\Hat{\sigma}_{3}\times F_{2}^{M}\Hat{\sigma}_{2}\bigg)+\bigg(\omega_{osc}\Hat{\sigma}_{3}\times F_{1}^{M}\Hat{\sigma}_{1}\bigg)\\
    & -2b\mathbf{F}+\bigg(e^{-i\pi/2}\frac{\eta}{2}\sin{\xi}\Hat{\sigma}_{2}\times F_{3}^{M}\Hat{\sigma}_{3}\\
    & +e^{-i\pi/2}\frac{\eta}{2}\sin{\xi}\Hat{\sigma}_{2}\times F_{1}^{M}\Hat{\sigma}_{1}\bigg)\\
    & =\mathbf{B}\times\mathbf{F}-2b\mathbf{F}+e^{-i\pi/2}(\mathbf{E}\times\mathbf{F})
\end{split}
\end{equation}
 %\begin{figure*}[htb!]
%\centering
%\includegraphics[scale=0.40]{decay 3.pdf}
%\includegraphics[scale=0.40]{decay 4.pdf}
%\includegraphics[scale=0.40]{decay 5.pdf}
%\vspace*{-3 cm}
%\caption{Plot of variation of eigenvalues with energy(GeV) in vacuum without parameterising the decay Hamiltonian}
%\label{plot:unifn_wo_threshold}
%\end{figure*}
\end{itemize}
Where we have denoted $\mathbf{E}=\frac{\eta}{2}\sin{\xi}\Hat{\sigma}_{2}+\frac{\eta}{2}\sin{\xi}\Hat{\sigma}_{2}$. This above equation depicts that in the presence of $\xi$, $\mathbf{F}$ not only tends to execute circular motion around $\mathbf{B}$, but also it does evolve around $\mathbf{E}$. This modifies the evolution equation 
\begin{equation}
\begin{split}
\Ddot{F}_{1}^{M}+2b\Dot{F}_{1}^{M}+\omega_{1}^{2}F_{1}^{M} &=0\\
    \Ddot{F}_{2}^{M}+2b\Dot{F}_{2}^{M}+\omega_{osc}^{2}F_{2}^{M} &=0\\
    \Ddot{F}_{3}^{M}+2b\Dot{F}_{3}+\eta\frac{\sin{\xi}}{2i}\Dot{F}_{1}^{M} &=0
\end{split}
\end{equation}
Where, $\omega_{1}=\sqrt{\omega_{osc}^{2}+b^{2}-\bigg(\frac{\eta^{2}\sin{\xi}^{2}}{4}\bigg)}$. This clearly shows that the presence of $\xi$ profoundly modifies the oscillation frequency, affecting the system as a perturbation introducing quantum fluctuation.
The relative dependence of the values of the parameters $\omega$ and $b$ guide us to probe the result in the following three categories.

  \textit{\textbf{1. Under damped oscillation}:} The most general case is when the oscillation parameter $\omega_{osc}$ takes over the damping parameter $b$, i.e., $\omega_{osc}>>\omega_{1}$. The larger value of the oscillation parameter than the damping parameter signifies the existence of both appearance and disappearance probability with an exponential decrease in amplitude. The amplitude of the flavor state gradually dies down. Consequently, $\mathbf{F}$ tries to align itself along the $\mathbf{B}$, which signifies that the flavor state exponentially damps down to the mass state. Geometrically, the phenomenon is depicted in the right panel of Fig.\ref{fig:3}. In the presence of the decay, the $\sigma_{3}^{M}$ axis is displaced by an amount b(red dashed line). The yellow dashed line denotes the system's total Hamiltonian, $\mathbf{B}$. The initial flavor state is aligned along $\sigma_{3}^{F}$, i.e., $F(0)$. The existence of decay enforces $\mathbf{F}$ to execute a spiral motion. It is to be mentioned here that for the critically damped case, the angular velocity modifies to $\sqrt{\omega^{2}_{osc}+b^{2}}$, although $\omega_{osc}>>b$. Moreover, this motion is not confined to the $\sigma_{2}\sigma_{1}$ plane. The appearance of $e^{-2bt}$ along the $\sigma_{3}$ drags $F(t)$ towards $\sigma_{3}$ compel the spiral rotation to follow a helical trajectory as shown in the figure.

 \textit{\textbf{2.Critically damped oscillation}:} In this scenario, the oscillation parameter and damping parameter, i.e., $\omega$ and $b$ merge together,i.e., $\omega_{osc}=b$. 
     The appearance probability exponentially damped down in this case. There is no oscillation of the flavor state associated with it. This implies that the flavor state does not execute the rotational motion around $\mathbf{B}$.
     With the decrement of the appearance probability, the disappearance probability increases and reaches a maximum, after which it reduces to zero. This implies that neutrinos tend to oscillate, but due to high perturbation, they damped down quickly to the mass basis. The diagram in the left panel of Fig.\ref{fig:3} interprets this phenomenon geometrically. Despite the fact that $\sqrt{\omega^{2}_{osc}+b^{2}}$, as the damping factor $2b$ equals the standard oscillation frequency $\omega$, the $\mathbf{F}$ damps down quickly to the mass basis. The figure suggests that $F(0)$ is set to rotate around $\mathbf{B}$. But due to the high damping coefficient, which merges with the standard oscillation frequency, the $\mathbf{F}$ can not complete one complete revolution and dies down quickly to the mass basis.

  \textit{\textbf{3. Overdamped oscillation}:} the 
     appearance probability decreases rapidly in this scheme. Initially, the neutrino tends to oscillate. Due to the negligible value of the disappearance probability, it can be neglected. Without spiraling in, the flavor state decays directly to the mass basis and loses its identity, as explained by Fig.\ref{fig:4}. It refers that due to the overtaking of the value of $b$ than $\omega$, $\mathbf{F}$ directly falls onto the mass basis. Without spiraling in.
      %\begin{figure*}[htb!]\label{fig:5}
  %\centering
%\includegraphics[scale=0.15]{over_app.pdf}
%\includegraphics[scale=0.15]{over_dis.pdf}
%\caption{Plot of variation of eigenvalues with energy(GeV) in vacuum without parameterising the decay Hamiltonian}
%\label{plot:unifn_wo_threshold}
%\end{figure*}   

From the above analysis, it is clear that the neutrino system behaves like a mechanical oscillator. Although the neutrino has been treated as an ultra-relativistic particle with quantum mechanical properties throughout the derivation, it starts to behave as a classical oscillator in the presence of an external perturbing Hamiltonian. For a mechanical system, the equation of motion in the presence of the viscous drag is $\Ddot{x}+2\gamma\Dot{x}+\omega_{0}x=0$, where $\gamma$ factor is associated with the damping of the system. In the case of the amplitude of displacement getting damped down, as in the case of the neutrino, the amplitude of the oscillation probability dies down. This implies that the presence of decay neutrino starts to mimic the mechanical oscillator.

 The above discussion encompasses the fact that the flavor state neutrino decays to its mass state, which causes the flavor of the neutrino to disappear after a certain distance completely, and the neutrino propagates as a flavorless massive particle. In the critically damped oscillation, the flavor state of the neutrino set off its journey towards accomplishing the circular trajectories. 
However, the marginal value of the decay parameter, which coincides with the decay parameter, restricts its motion, and they could not complete one complete revolution. Consequently, there is some finite probability of disappearance occurring, although the oscillation length can not be achieved, and the trajectory spirals towards the mass basis similar to the Fibonacci sequence \cite{Sudipta_2019yyu}. Fig.\ref{fig:3} depicts the same phenomenology as described. For the over-damped oscillation, due to the substantial value of the decay parameter $b$, i.e., greater than the standard oscillation, the flavor state directly dies down to the mass basis without performing oscillation.  It is also to be mentioned that the system of the decaying flavor state of neutrino is precisely similar to that of a nuclear spin state in an external Magnetic \cite{Am-Shallem_2015}. This also indicates the non-adiabaticity of the flavor state oscillation in the appearance of a perturbing Hamiltonian in the neutrino system. However, in the probability regime, we have considered the adiabatic transition. But figuratively, there is a clear-cut indication that the mixing angle should also vary in the presence of decay. The flavor vacuum oscillation of the neutrino indicates that the flavor state picked up from the vacuum during propagation, and the initial flavor state of the neutrino acts as a coupled harmonic oscillator.
%This also leads to the parallel phenomenological explanation of the solar neutrino problem. 
\newline
%\red{Paragraph on CP violations in presence of decay}
%\blue
We find that the appearance probability and disappearance probabilities in the presence of decay are dependent on phase $\zeta$. The Majorana phase($\phi$) also appears in the probability analysis if the off-diagonal decay terms are present. Thus, we have two phases $\zeta$ and $\phi$
which can induce CP violation in a vacuum. The anti-neutrino oscillation probabilities can be easily obtained by replacing $\zeta \rightarrow -\zeta$ and $\phi \rightarrow -\phi$. To quantify the amount of CP violation associated with the presence of decay, we define the quantity $A^{CP}_{\alpha \beta} = P^{decay}_{\alpha \beta} - \overline{P}^{decay}_{\alpha \beta}$. The asymmetry associated with muon appearance probability, $A^{CP}_{e\mu}$ is proportional to factor $e^{-2bt} \sin(\zeta -\phi)B$, where $ B = \frac{\sin{2\theta} \sin^2 ({\Delta m^2 L/4E})}{\Delta m^2/2E}$\cite{Dixit:2022izn}.
%\blue

%%%%%%%%%%%%%%%%%%%%%%%%%%%%%%%%%%%%%%%%%%%%%%%%%%%%%%%%%%%%%%%%%%%%%
%%%%%%%%%%%%%%%%%%%%%%%%%%%%%%%%%%%%%%%%%%%%%%%%%%%%%%%%%%%%%%%%%%%%%
\section{Comparison with neutrino oscillation in vacuum and matter}

 We provide an overview of the geometric representation of neutrino oscillations in both vacuum and in the presence of matter, as previously explored in references \cite{Kim:1987bv, Kim:1993byl, Giunti:2007ry}. Subsequently, we compare these findings with our current analysis, which incorporates the geometric representation of neutrino oscillations along with decay effects.%blue
\subsection{Description of neutrino oscillations in vacuum}
Before going to our results for geometrical interpretation in the presence of decay, we provide an overview of the geometrical interpretation in vacuum and matter explored in the literature.

Let's briefly outline the geometric representation of two-flavor neutrino oscillations, as initially discussed by Kim et al. (1988) \cite{Kim:1987bv, Sen:2018put}. This illustration utilizes the evolution equation for two generations of neutrinos. Using orthonormal unit vectors, the authors explained the geometrical picture of neutrino oscillations in vacuum and matter. The propagation and time evolution of neutrino flavor states are described as follows:

	\begin{eqnarray}
i\frac{\partial }{\partial t}\left[
	\begin{array}{c}
	 \nu_e \\
	 \nu_\mu
	\end{array}
	\right]&=&H_{\rm F}\left[
	\begin{array}{c}
	 \nu_e \\
	 \nu_\mu
	\end{array}
	\right]
= - \frac{1}{2} \overrightarrow{\sigma} \cdot \overrightarrow{B} 
\left[
	\begin{array}{c}
	 \nu_e \\
	 \nu_\mu
	\end{array}
	\right]
	\end{eqnarray}
	
where $$\overrightarrow{B} = \frac{1}{2E} \bigg[-\Delta m^2 \sin 2\theta\, \hat{x}
     + (\Delta m^2 \cos 2\theta - A) \, \hat{z}
      \bigg]$$
here $A=2\sqrt{2}G_F N_e E$ with $G_F$ Fermi constant, $N_e$ the electron number density in matter, and $E$ neutrino energy. 
The vector B in the presence of matter is linked to its value $B_0$ in vacuum ($A=0$) as $$B = B_0-(A/2E)\hat{z}  $$

Defining $\psi \psi^\dagger = \frac{1}{2} \big[\mathbb{1} + \overrightarrow{\sigma} \cdot \overrightarrow{P}\big]$, where $\overrightarrow{P}$ represents the equivalent magnetic moment vector (or polarization vector), the pure electron neutrino state $\nu_e$ can be visually depicted with $\overrightarrow{P} = \hat{z}$, while the pure muon neutrino state $\nu_\mu$ is oriented along the negative z-axis ($\overrightarrow{P} = -\hat{z}$). The concept of neutrino oscillation can be interpreted as the precession of the magnetic moment vector ($\overrightarrow{P}$) about an external magnetic field $\overrightarrow{B}$.
	\begin{equation}
 \frac{d \overrightarrow{P}}{d t} = \overrightarrow{P} \times \overrightarrow{B}\,.
\end{equation}

\begin{figure}[htb!]
% \centering
\includegraphics[scale=0.18]
{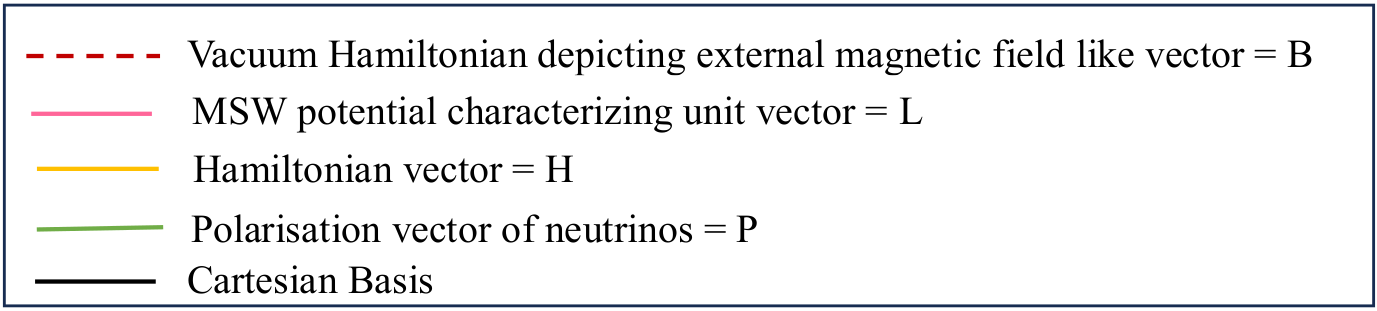}
\includegraphics[scale=0.42]{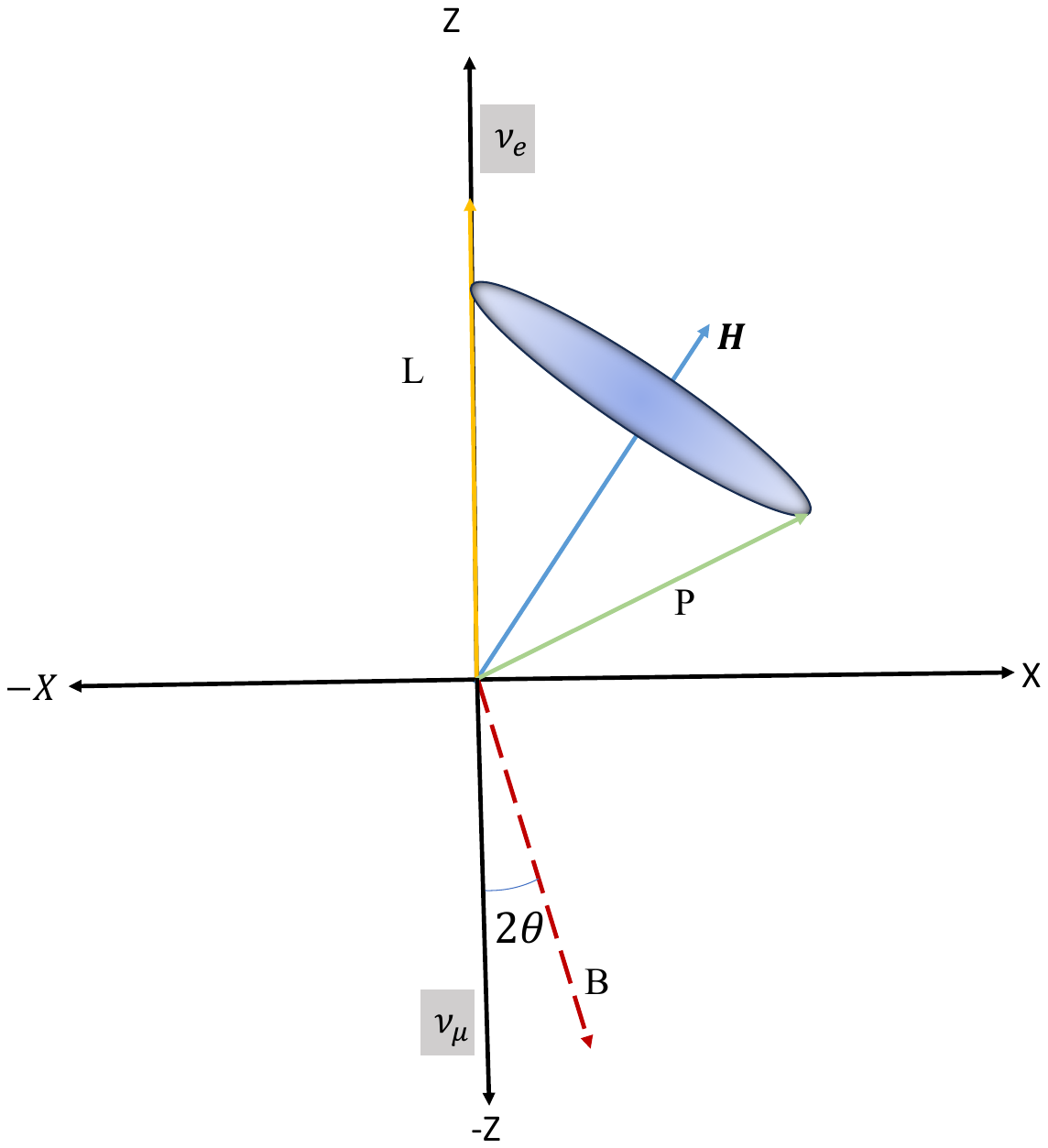}
\vspace{-3.5cm}
\caption{Illustration of the geometrical interpretation of two flavor neutrino oscillations in vacuum and matter effect. The choice of orthonormal basis is in the Cartesian coordinate system where the vectors $\Vec{\sigma}$, $\Vec{B}$, $\Vec{P}\equiv \Vec{F}$ etc. are derived. This is the picturization of the precession of the polarisation vector or neutrino flavor state around the effective Hamiltonian in vacuum as well as in matter. Initially, the pure electron-neutrino state is defined as $\Vec{P}(t=0)$ along the direction of the z-axis. The evolution of this flavor state can be understood by the quantity $\Vec{H}\times\Vec{P}$, where $\Vec{H}$ is defined for effective Hamiltonian for vacuum or matter. The other muon-neutrino flavor state is depicted in the -z axis. \textit{Kim.et.al (1988)}~\cite{Kim:1987bv, Sen:2018put} }
\label{fig:5}
\end{figure}

B is a constant vector ($B_0$) in vacuum, and P describes the cone's surface with axis B and an opening angle 2$\theta$ as shown in Fig.\ref{fig:5}. 
%blue
%blue
Within the core of the Sun, we have mainly neutrino associated with electrons such that $\overrightarrow{P}$ is directed along the z-axis. Also, due to high electron density for the propagating neutrinos, $\overrightarrow{B}$ is almost parallel to the negative z-axis, so the precession cone is close to $180^\circ$. One can visualize it as if P is rotating about $-B$ in an anticlockwise direction with an angle close to zero. When the neutrinos emerge from the Sun, the effective matter potential decreases, and $\overrightarrow{B}$ shifts to its vacuum value $\overrightarrow{B_0}$. P will precise around $\overrightarrow{B_0}$ with opening angle $180^\circ$ for adiabatic migration. Thus, in the final state, we have a neutrino state dominated by $\nu_{\mu}$ explaining the geometrical picture of the MSW effect inside the Sun.
%blue
%blue
 The work carried out by Kim et al. was mainly focused on one initial neutrino flavor. However, it is useful to understand the geometrical representation for an ensemble of neutrinos with different flavors.  We can define such incoherent neutrino beam by using density matrix formalism used by \cite{Giunti:2007ry}.
%blue 
A Hermitian density operator can define the neutrino beam. 
\begin{equation}
 \rho(x) = \sum_{\alpha} |\nu_\alpha(x)\rangle\, W_\alpha \, \langle \nu_\alpha(x) |
 \label{eq:comp-vac-1}
\end{equation}
The parameter $W_\alpha$ is defined as the initial statistical weight factor, which is a measure of the probability of particular neutrino flavor $\alpha$ at $x=t=0$. 
The key property of density matrix operator is $\mbox{Tr}\big[\rho(x) \big] = \sum_{a} \langle \nu_a(x) |\, \rho(x) \, | \nu_a(x) | =1$ where $\big\{|\nu_a\rangle \big\}$ is a complete set of states. The evolution equation of the density matrix in the flavor basis is derived in \cite{Giunti:2007ry} and is given by
\begin{eqnarray}
 i \frac{d \rho_{\rm F}}{d x} = \big[H_{\rm F}, \rho_{\rm F} \big] 
\label{eq:comp-vac-2}
\end{eqnarray}
\begin{figure*}[htb!]
% \centering
\includegraphics[scale=0.35]{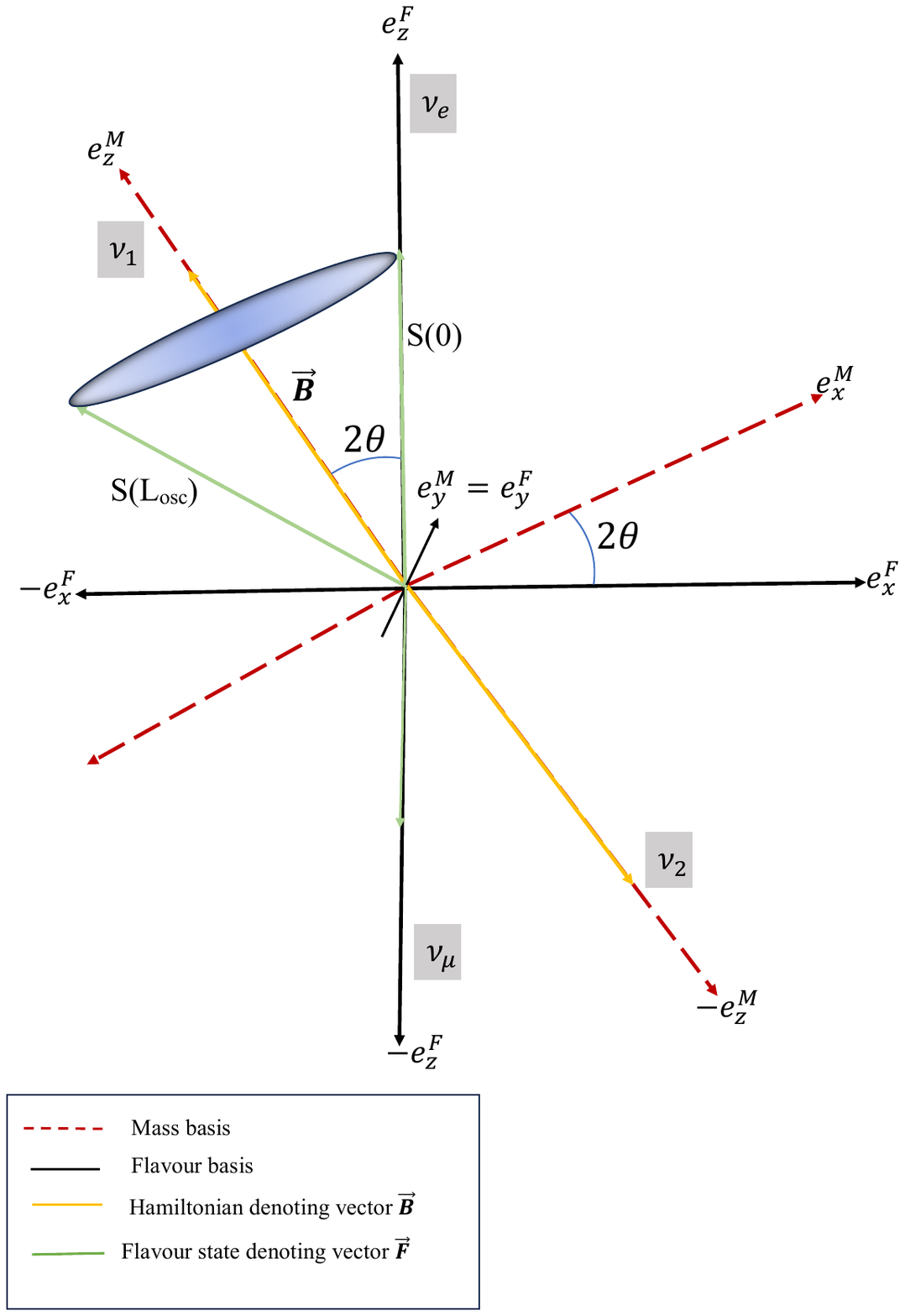}
\hspace{0.5 cm}
\includegraphics[scale=0.35]{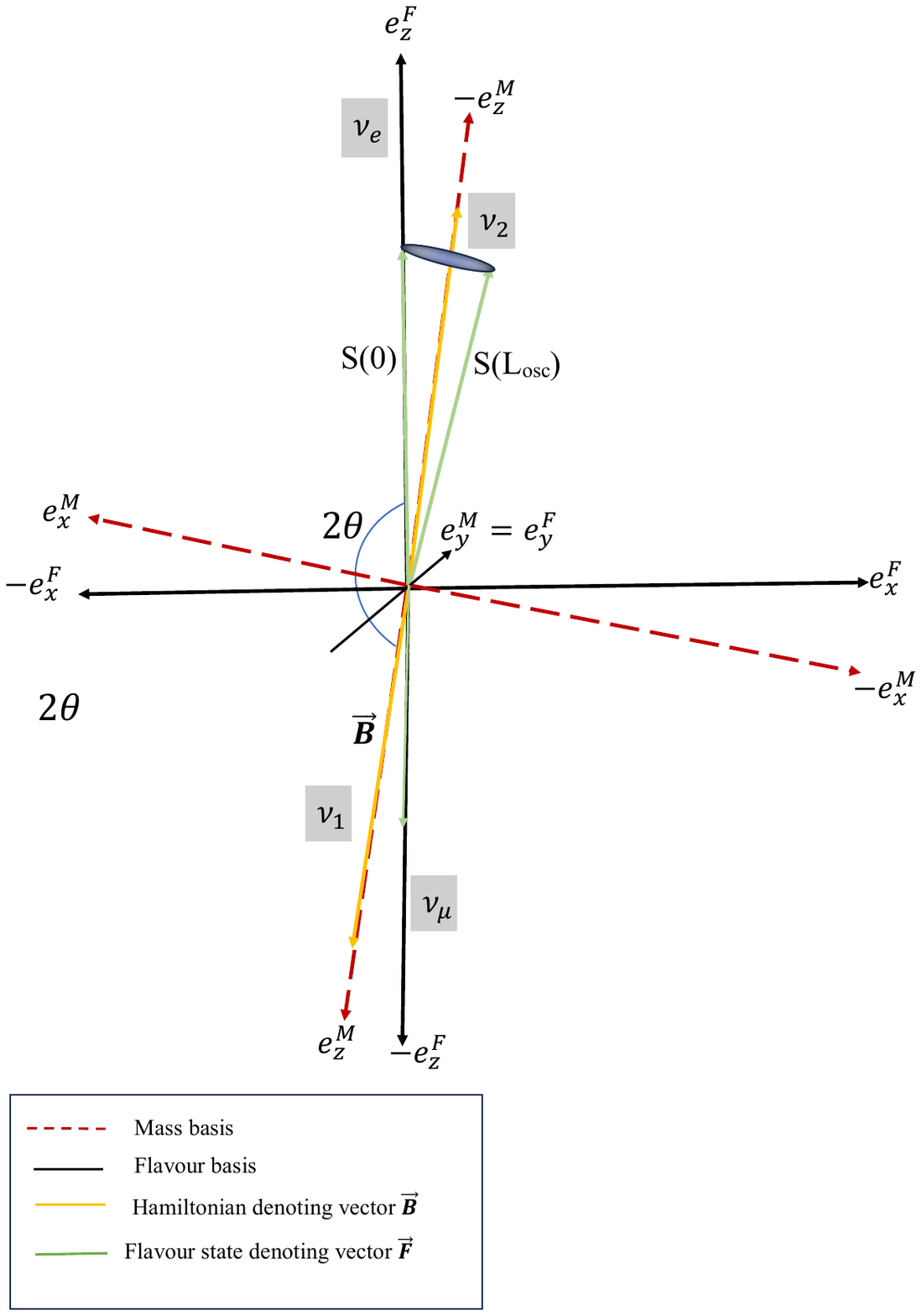}
\vspace*{0 cm}
\caption{Geometrical representation of neutrino oscillations in vacuum (in left-panel) and in presence of matter (right-panel). The choice of basis adopted here is the flavor basis of neutrinos with the three orthonormal vectors  $\vec{e}_{1F}$, $\vec{e}_{2F}$ and $\vec{e}_{3F}$ represented by black solid lines.The mass eigenstates of neutrinos are derived from equivalent orthonormal vectors represented by dashed magenta color lines and are related to flavor ones by mixing angle $2\theta$. The effective Hamiltonian in two neutrino systems in a vacuum is described in flavor basis by $H_{F} = - 1/2\,\vec{\sigma}_{F}$ while the vector $\vec{B} = -\frac{\Delta m^2\, \sin 2\theta}{2 E}\, \vec{e}_{1F} + \frac{\Delta m^2\, \cos 2\theta}{2 E} \vec{e}_{3F} $ is depicted in the second quadrant. The initial neutrino flavor state, let us say $\nu_e$, is oriented along $\vec{e}_{3F}$ direction and is represented by a flavor state vector $\vec{S}(t=0)$. The evolution of the neutrino state vector $\vec{S}$ from distance $x=0$ to $x=L_{osc}$ is described by the quantity $\vec{S} \times \vec{B}$ and thereby, the precision of $\vec{S}$ about $\vec{B}$ describes neutrino oscillation.}
\label{fig:6}
\end{figure*}
with initial boundary condition $[\rho_{\rm F}]_{rs} (0) = W_r \delta_{rs}$. With two flavor scenario with $\nu_e-\nu_\mu$, the effective Hamiltonian $H_{\rm F}$ in flavor basis can be written down as
\begin{equation}
 H_{\rm F} = -\frac{1}{2} \overrightarrow{\sigma_{\rm F}} \cdot \overrightarrow{\text{B}}\quad \,
            \rho_{\rm F} = \frac{1}{2} \big[\sigma_{0} + 
            \overrightarrow{\sigma_{\rm F}} \cdot \overrightarrow{P}\big]\,. 
\label{eq:comp-vac-3} 
\end{equation}
Here, the choice of basis is essential and as adopted in \cite{Giunti:2007ry}, $\overrightarrow{e}_{1F}$, $\overrightarrow{e}_{2F}$, $\overrightarrow{e}_{3F}$ are three orthonormal basis vectors with property $\overrightarrow{e}_{aF} \cdot \overrightarrow{e}_{bF} =\delta_{ab}$ in which flavor states of neutrinos can be constructed. 

The components of vectors $\overrightarrow{B}$ (can be interpreted as a magnetic field) defined in the flavor basis are 
\begin{eqnarray}
 &&B_{1F} = -\frac{\Delta m^2\, \sin 2\theta}{2 E}\,\nonumber \\
 &&B_{2F} = 0\,\nonumber \\
 &&B_{3F} = \frac{\Delta m^2\, \cos 2\theta}{2 E}\,.
\end{eqnarray}
%blue
For the evolution of the system in the presence of matter, the mass square difference($\Delta m^2$) and the mixing angle($\theta$) should be replaced by the effective mass square difference( $\Delta \Tilde{m}^2$) and modified mixing angle ($\theta_M$) in the presence of matter potential.
%blue
The components for equivalently interpreted Polarisation vector ($\overrightarrow{P}$) arising from the density matrix operator as,
\begin{eqnarray}
 &&P_{1F} = 2 \mbox{Re} [\rho_{\rm F}]_{e\mu}\,\nonumber \\
 &&P_{2F} = - 2 \mbox{Im} [\rho_{\rm F}]_{e\mu}\,\nonumber \\
 &&P_{3F} = [\rho_{\rm F}]_{ee} - [\rho_{\rm F}]_{\mu\mu} \,.
\end{eqnarray}
%blue 
With the orthonormal set of basis vectors $\overrightarrow{e}_{a}$, the various vectors defined as $\overrightarrow{\sigma}_{\rm F} = \sum^{3}_{a=1} \sigma^{a} \overrightarrow{e}^{a}_{\rm F}$, $\overrightarrow{B} = \sum^{3}_{a=1} B^{a}_{\rm F} \overrightarrow{e}^{a}_{\rm F}$, and $\overrightarrow{P} = \sum^{3}_{a=1} P^{a}_{\rm F} \overrightarrow{e}^{a}_{\rm F}$.
They derived from evolution equations of the density matrix operator, and the corresponding translated initial boundary conditions are $\overrightarrow{P}_{1F}(0)=0$, $\overrightarrow{P}_{2F}(0) = 0$ and $\overrightarrow{P}_{3F}(0) = W_e-W_\mu$. 

Let us understand the physical meaning of neutrino oscillations in vacuum in terms of $\overrightarrow{\sigma}$, $\overrightarrow{B}$ and $\overrightarrow{P}$.

Assume that the initial neutrino state is a pure electron neutrino produced in Sun, i.e., $W_e=1$ and $W_\mu=0$. The Polarisation vector $\overrightarrow{P}(0)$ is aligned along $\overrightarrow{e}_{3F}$ direction. For pure muon type neutrino state ($W_e=0$ and $W_\mu=1$) the Polarisation vector $\overrightarrow{P}(0)$ will be aligned along negative $\overrightarrow{e}_{3F}$ direction. For an incoherent mixture of electron and muon neutrino states, the orientation of the Polarisation vector $\overrightarrow{P}(0)$ will be along $\pm \overrightarrow{e}_{3F}$ depending upon the relative strength of $W_{r}$ with $r=e,\mu$. The norm of the Polarization vector will decide whether the neutrino beam is a pure state or incoherent mixture.
%blue 
The precision of $\overrightarrow{P}$ around the $\overrightarrow{B}$ can have different geometrical viewpoints depending upon the value of matter density compared to the resonance density. We present one such scenario in Fig.\ref{fig:6}, where
the transition is adiabatic for solar neutrinos.
%blue
The right panel of Fig.\ref{fig:6} suggests the geometrical interpretation of neutrino oscillation in matter in over $e^{F}_{i}$ and $e^{M}_{i}$ as described in \cite{Giunti:2007ry}. The presence of matter Hamiltonian $A=2\sqrt{2}G_{F}N_{e}E$, the standard Hamiltonian of the neutrino is dragged down as the figure indicates. However, the  $\Vec{P}$ tends to follow the precessional motion around $\Vec{B}$. In order to retain its circular motion $\Vec{P}$ starts to revolves round $-\Vec{B}$ as a results of which instead of $\nu_{1}$ it is now coincide with $\nu_{2}$ and $\Vec{P}$ describes a narrow cone $(\sim 180^{\circ})$ round $\Vec{B}$. The resonance crossing of $\Vec{P}$ in the nonadiabatic case represents $\nu_{2}\rightarrow \nu_{1}$\cite{Giunti:2007ry} 
We find that both the works by Kim and Guinti describe neutrino oscillations in vacuum and matter but with different basis choices. The choice of orthogonal basis is not unique. For example, Mikheyev and Smirnov \cite{Mikheev:1987jp} have described the two-flavor neutrino oscillation using an orthogonal mass eigenstates as the choice of basis as $\nu_1$, $Re~\nu_2$, and $Im~ \nu_2$. Nevertheless, in the following subsection, we present the geometrical interpretation of neutrino oscillations in the presence of decay, which is not yet clearly explored in the literature.
%blue
\subsection{Description of neutrino oscillation with decay}
The standard Hamiltonian of the neutrino oscillation is defined as $\mathcal{H}_{0}=\begin{pmatrix}
        \frac{m^{2}_{1}}{2E} & 0\\
        0 & \frac{m^{2}_{2}}{2E}\\
    \end{pmatrix}$. Following\cite{Dixit:2022izn} this Hamiltonian can be decomposed as $\mathcal{H}_{0}=\frac{1}{2}(\omega_{0}\mathbf{I}+\omega\cdot\sigma_{z})$ with $\omega_{0}=\frac{(m^{2}_{1}+m^{2}_{2})}{2E}$ and $\omega=\frac{(m^{2}_{1}-m^{2}_{2})}{2E}$. Since the $\omega_{0}$ term doesn't contribute to the flavor-changing neutrino oscillation, the first term of the $\mathcal{H}_{M}$ is ruled out. As we have already discussed, we can unify the effect of invisible neutrino decay by the Hamiltonian,
    \begin{equation}
    \begin{split}
        \mathcal{H}_{d} & =\frac{-i}{2}\begin{pmatrix}
            b & \eta/2\\
            \eta/2 & b\\
        \end{pmatrix}=\frac{-i}{2}(b\mathbf{I}+\eta\cdot\sigma)
    \end{split}
    \end{equation}
Which modifies the description of the Hamiltonian of the neutrino system to $\mathcal{H}_{m}=\frac{1}{2}(-ib\mathbf{I}+\omega\cdot\sigma-i\eta/2\cdot\sigma)$. 
This results in a modification of the neutrino system's Hamiltonian to $\mathcal{H}_{m}=\frac{1}{2}(-ib\mathbf{I}+\omega\cdot\sigma-i\eta/2\cdot\sigma)$. In summary, when considering the presence of decay, the Hamiltonian for the neutrino system can be vectorized by introducing a vector that represents decay in addition to the standard Hamiltonian for neutrino oscillation, denoted as $\mathcal{H}_{m}=\mathcal{H}_{0}+\mathcal{H}_{d}$. This is mathematically elucidated in the following equation.
%blue
\begin{equation}
\begin{split}
    \mathcal{H}_{0} & =\frac{1}{2}(\omega_{0}\mathbf{I}+\omega\cdot\sigma_{3})\\
    \mathcal{H}_{d} & =\frac{-i}{2}(b\mathbf{I}+\eta\cdot\sigma_{1})\\
\end{split}
\end{equation}
%To represent it geometrically, we have chosen the Pauli matrix basis. The motivation for this choice is that the Pauli matrices can be expanded over the mass basis, which can be thought of as a two-level system with two mass eigenstates $\ket{\nu_{1}}$ and $\ket{\nu_{2}}$. Explicit form of $\sigma_{i}(i=x,y,z)$ over the mass basis is given by \ref{eq:12}. The distinctive feature of the superposition of two non-degenerate mass eigenstates over the flavor eigenstate of the neutrino system encompasses us for adopting such formalism, which is quite different for other particles where a unique mass eigenstate describes each flavor eigenstate. Such choice of basis vectorized the Hamiltonian on the mass basis with the basis vector $\Hat{\sigma}_{i}$. In the mass basis, we have denoted the vector representation of the Hamiltonian for the standard oscillation by $\mathbf{B}_{h}$ and decay Hamiltonian $\mathbf{B}_{nh}$.
%blue
For a geometric representation, we have opted for the Pauli matrix basis. This choice is motivated by the ability to expand the Pauli matrices over the mass basis, which can be conceptualized as a two-level system featuring two mass eigenstates, namely $\ket{\nu_{1}}$ and $\ket{\nu_{2}}$. The explicit form of $\sigma_{i}$ (where $i=1,2,3$) over the mass basis is provided in eq.(\ref{eq:12}). The unique aspect of the superposition of two non-degenerate mass eigenstates over the flavor eigenstate of the neutrino system motivates the adoption of this formalism. This contrasts with other particles, where a unique mass eigenstate exclusively describes each flavor eigenstate. The selection of this basis allows for the vectorization of the Hamiltonian in the mass basis, represented by the basis vector $\Hat{\sigma}{i}$. In the mass basis, we denote the vector representation of the Hamiltonian for standard oscillation as $\mathbf{B}_{h}$ and the decay Hamiltonian as $\mathbf{B}_{nh}$.
%blue
\begin{equation}
\begin{split}
    \mathbf{B}_{h} & =\frac{1}{2}\omega\Hat{\sigma}_{3}\\
    \mathbf{B}_{nh} & =\frac{-i}{2}(b\mathbf{I}+\eta/2\Hat{\sigma}_{1})\\
    \mathbf{B} & =\frac{1}{2}(\omega\Hat{\sigma}_{3}+e^{\frac{-i\pi}{2}}(b\mathbf{I}+\eta
    \Hat{\sigma}_{1}))
\end{split}
\end{equation}
The imaginary $-i$ factor physically represents the rotation of the coordinate axes by the angle of $\frac{-i\pi}{2}$. After elucidating the vectorized Hamiltonian of the neutrino system, we have introduced the density matrix formalism to express the flavor states. Similarly, as we have denoted the Hamiltonian in vector representation, we have demonstrated the density operator of the neutrinos depicting the flavor state as,
\begin{eqnarray}
\rho=\mathbf{F}\cdot\sigma=F_{1}\Hat{\sigma}_{1}+F_{2}\hat{\sigma}_{2}+F_{3}\Hat{\sigma}_{3}
\end{eqnarray}
The Magnitude of the $\mathbf{F}$ is $F_{i}=\mbox{Tr}\frac{1}{2}\sigma_{i}\rho$.
For neutrino system, $\rho(t=0)=\begin{pmatrix}
    \cos^{2}{\theta} & \cos{\theta}\sin{\theta}\\
    \cos{\theta}\sin{\theta} & \sin^{2}{\theta}\\
\end{pmatrix}=\begin{pmatrix}
    F_{3} & F_{1}-iF_{2}\\
    F_{1}+iF_{2} & -F_{3}\\
\end{pmatrix}$
The interpretation directly shows $\mathbf{F}$ in terms of the component of $\rho$ as,
\begin{eqnarray}
\begin{split}
    F_{1} & =(\rho_{12}+\rho_{21})\\
    F_{2} & =i(\rho_{12}-\rho_{21})\\
    F_{3} & =(\rho_{11}-\rho_{22})
\end{split}
\end{eqnarray}
Thus, two operators, Hamiltonian, and density operator, are represented by two vectors $\mathbf{B}$ and $\mathbf{F}$ respectively. The dynamical evolution equation of $\mathbf{F}$ describes how these two vectors are interrelated. 
In this demonstration, instead of denoting the Hamiltonian as an inner product with the Pauli matrices on some arbitrary basis, we have expressed the vectorized Hamiltonian in Pauli bases. In this context, it is to be mentioned that the main reason for introducing the Pauli basis is that during the change of basis from mass to flavor or \textit{vice-versa} only these basis vector $\Hat{\sigma}_{i}$ will change which makes the visualization of the phenomenon of oscillation and mathematics much simpler. This flavor-mass bases conversion is given as,
\begin{equation}\label{eq:41}
\begin{split}
    \sigma_{0}^{M} &=\sigma_{0}^{F}\\ \sigma_{1}^{M} & =\sin{2\theta}\sigma_{3}^{F}+\cos{2\theta}\sigma_{1}^{F}\\
    \sigma_{2}^{M} & =\sigma_{2}^{F}\\
    \sigma_{3}^{M} & =\cos{2\theta}\sigma_{3}^{F}-\sin{2\theta}\sigma_{1}^{F}\\
\end{split}
\end{equation}
The scenario also addresses that the Hamiltonian is responsible for the evolution of flavor state with either time. Following the Von-Neumann equation, the evolution equation of the density operator is explicitly of the form,
\begin{equation}\label{eq:49}
    i\frac{d}{dt}\rho(t)=\frac{-i}{2}[\mathcal{H}_{0},\rho]-\frac{1}{2}\{\mathcal{H}_{d},\rho\}
\end{equation}
This follows from the Schrodinger equation, which gives the dynamical evolution of the operator. In the present representation, this equation is written as $\frac{d}{dt}F_{i}=\mbox{Tr}\frac{1}{2}\sigma_{i}\frac{d}{dt}\rho$ which gives the dynamical flavor evolution over Pauli basis in terms of state vector as,
\begin{equation}\label{eq:50}
    \frac{d}{dt}\mathbf{F}=\mathbf{F}\times\mathbf{B}-2b\mathbf{F}
\end{equation}
In this framework, the dependency of the evolution of the state for the two-flavor neutrino oscillation over the off-diagonal terms of the decay Hamiltonian is automatically canceled out. 
\begin{figure}[htb!]
 \centering
 \hspace*{-1cm}
\includegraphics[scale=0.36]{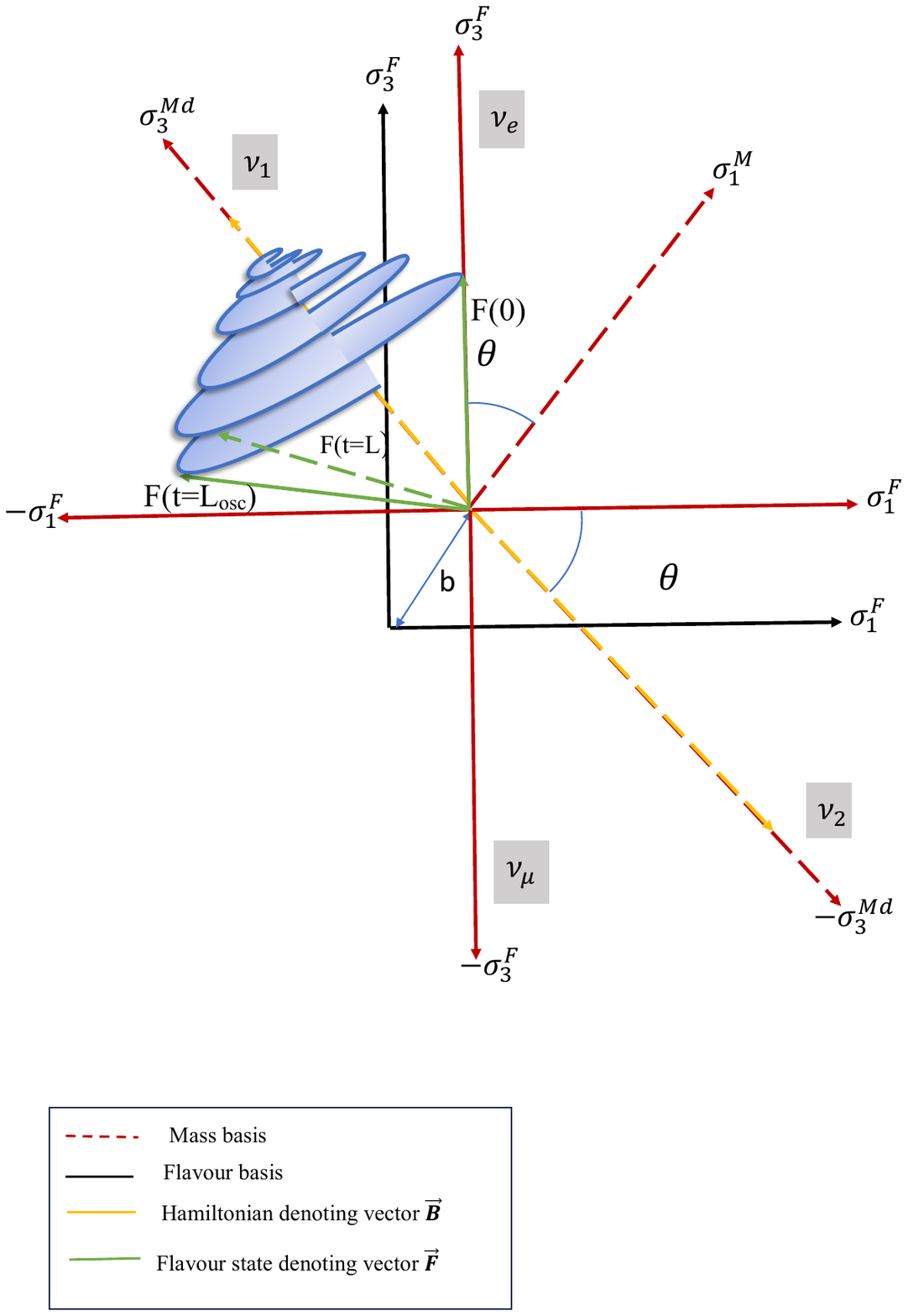}
\vspace*{0 cm}
\caption{Figures illustrate the geometrical representation of neutrino oscillation in decay under damped case in Pauli basis. For the under-damped case, the oscillation amplitude damped down in an oscillatory way. Consequently, there is oscillation with decreasing amplitude, which directs towards the fact that after one full revolution, it is impossible to get back to the initial flavor state as the presence of decay parameter $b$, which causes the flavor state depicting vector $\mathbf{F}$ execute a helical motion. Moreover, this helical motion doesn't confine to a plane because this decay parameter affects the $\Hat{\sigma}_{3}$ components as well, which makes $\mathbf{F}$ drag toward $\Hat{\sigma}_{3}^{Md}$ which the mass basis in the presence of decay. In addition, with the gradual decrement of the pure state, the flavor state dies down, demonstrating that it ultimately dissipates to the mass state and $\mathbf{F}$ aligns itself towards $\mathbf{B}$.}
\label{fig.7}
\end{figure}
It is also to be mentioned here that the flavor state representing vector $\mathbf{F}$ is defined over the mass basis. At an initial time $t=0$, we have considered the neutrinos are produced in the pure electron flavor state. Therefore, it is clear from eq.(\ref{eq:19}) that solving the equation and transforming it to the flavor basis following the change of bases from eq.(\ref{eq:41}) give the appearance and the disappearance probabilities. 
\begin{equation}\label{eq:51}
\begin{split}
    P_{e\mu} & =e^{-2bt}P_{e\mu}^{vac}\\
    P_{ee} & =e^{-2bt}P_{ee}^{vac}\\
\end{split}
\end{equation}
Where $P_{e\mu}^{vac}=\sin^{2}{2\theta}\sin^{2}{\omega L}$. 
The geometrical description of the eq.(\ref{eq:50}) is given in Fig.\ref{fig.7}, which implements the damped oscillatory nature of the neutrino flavor state. Figuratively, it suggests that the flavor eigenstate of neutrino decays to the mass eigenstates following helical trajectory due to the decay factor $e^{-2bt}$.
The work's uniqueness lies in the choice, simple mathematical description, and effective geometrical visualization. 

This illustration also shows that the neutrino system mimics the NMR system in the context of the similar format of the Bloch equation. From \cite{Am-Shallem_2015}, the Bloch equation for an NMR system in an external magnetic field is
\begin{equation}\label{eq:45}
\begin{pmatrix}
    \Dot{S}_{x}\\
    \Dot{S}_{y}\\
    \Dot{S}_{z}\\
\end{pmatrix} =\begin{pmatrix}
    -\frac{1}{T_{2}} & \Delta & 0\\
    -\Delta & -\frac{1}{T_{2}} & \epsilon\\
    0 & -\epsilon & -\frac{1}{T_{2}}\\
\end{pmatrix}\begin{pmatrix}
    S_{x}\\
    S_{y}\\
    S_{z}\\
\end{pmatrix}+\begin{pmatrix}
    0\\
    0\\
    S_{z}^{0}\\
\end{pmatrix}
\end{equation}
Where the Hamiltonian of the system, including the perturbation, is given by $\Hat{\mathbf{H}}=\Delta\Hat{\mathbf{S}}_{z}+\epsilon\mathbf{S}_{x}$. $S_{x}$, $S_{y}$, $S_{z}$ denotes the nuclear spin under the influence of an external magnetic field with $T_{1}$ and $T_{2}$ as the dissipation and dephasing relaxation parameters. 
Similarly, for the Bloch equation for the neutrino system as derived in eq.(\ref{eq:20}),
\begin{equation}\label{eq:45}
\begin{pmatrix}
    \Dot{F}_{x}\\
    \Dot{F}_{y}\\
    \Dot{F}_{z}\\
\end{pmatrix} =\begin{pmatrix}
    -b & -\omega & 0\\
    \omega & -b & 0\\
    0 & 0 & -2b\\
\end{pmatrix}\begin{pmatrix}
    F_{x}\\
    F_{y}\\
    F_{z}\\
\end{pmatrix}
\end{equation}
Where the role of the external magnetic field is played by the mass square differences $\omega$ and the dissipation parameter is $b$. Although the $x$ component of the Hamiltonian is present in both cases, for NMR, it is Hermitian; hence, it is present, $\epsilon$, and for neutrino it is non-Hermitian hence is doesn't show up in the Bloch matrix. However, in the Liouville super-operator formalism it will be present in the matrix defining the Liouville super-operator \cite{Shafaq:2021lju}. 
%%%%%%%%%%%%%%%%%%%%%%%%%%%%%%%%%%%%%%%%%%%%%%%%%%%%%%%%%%%%%%%%%%%%%%%%
%%%%%%%%%%%%%%%%%%%%%%%%%%%%%%%%%%%%%%%%%%%%%%%%%%%%%%%%%%%%%%%%%%%%%%%%%
\section{Conclusion}

We discussed the geometrical interpretation of two-flavor neutrino oscillation in the presence of neutrino decay with a detailed analysis of the theoretical framework. A comparative study of the geometrical description of neutrino oscillation performed in the present work along with earlier works~\cite{Kim:1987bv, Kim:1993byl, Giunti:2007ry} has been presented for better understanding of the reader. We followed the density matrix formalism to define the neutrino as an open quantum system in the presence of decay.
 We vectorized the operators, the Hamiltonian, and the density operator of the system to analyze the neutrino oscillation, including decay geometrically. Implementing such formalism, we devised a simple mathematical analysis following \cite{Am-Shallem_2015} to produce the expression of probabilities. Using the Pauli basis, we interpret the results schematically, which validates the decay of the flavor state to the mass state. we also showed the modified evolution of the flavor state depicting vector in presence of phase factor $\xi$ in the non-hermitian Hamiltonian. This introduced a $CP$ violation phase in two flavor neutrino oscillation.
 We also qualitatively discussed the three scenarios of under-damped, over-damped, and critically-damped motion. To give the work an overall completeness, we did a comparative study of the geometrical interpretation of neutrino oscillation in vacuum, matter, and decay that includes the MSW effect. In a nutshell, we have found that in the presence of decay, the flavor state depicting vector $\mathbf{F}$ executes a spiral motion around the Hamiltonian depicting vector $\mathbf{B}$, through which the flavor state of neutrino decays to the mass eigenstate. The choice of Pauli basis has made the interpretation of the phenomenon of decay in a more convenient way. However, such a choice of basis has made the oscillation probability independent of the off-diagonal factor, which stipulates the event of amplitude damping of the oscillation in addition to introducing a perturbation to the mass eigenstate.
 
The geometrical interpretation of two-flavor neutrino oscillation provides an easier way to understand why such oscillations get enhanced in matter- an effect known as MSW (Mikheyev-Smirnov-Wolfenstein) effect that could explain the well-known solar neutrino puzzle. This description can also apply to understanding the governing dynamics of Supernovae and the fast flavor conversion of Supernovae neutrinos \cite{Dasgupta:2018ulw, Glas:2019ijo, Sen:2017ogt}. The three-flavor neutrino oscillations can be approximated with two-flavor neutrino oscillations ($\nu_e$ and $\nu_x$ with $x=\mu, \tau$) within Supernovae as the muon and tau neutrinos have identical interactions and thereby, behaves similarly \cite{Bhattacharyya:2020dhu, Bhattacharyya:2020jpj}. Additionally, one may consider neutrino as an open quantum system with a similar viewpoint to the nuclear magnetic resonance (NMR) system \cite{Am-Shallem_2015}, where the external magnetic field acts as a beam splitter. In contrast, in the case of neutrino, the unitary transformation matrix performs the same. In addition to this, the proper characterization of the decay parameter with its value and its consequence on neutrino oscillation is still a broad area to investigate. Furthermore, the investigation of the proper reason behind the neutrino decay will make the understanding of neutrino oscillation more effective.   
\vspace{1cm}
%%%%%%%%%%%%%%%%%%%%%%%%%%%%%%%
\acknowledgments 
Rajrupa Banerjee would like to thank the Ministry of Electronics and IT for the financial support through the Visvesvaraya fellowship scheme for
carrying out this research work. RB is very thankful to Prof. Prasant. K. Panigrahi for the
fruitful discussion carried from time to time for the betterment of this work.
KS acknowledges the financial support received via the "Bi-nationally Supervised Doctoral Degree Program" from the German Academic Exchange Service (DAAD). Additionally, KS extends thanks to the  "Karlsruhe School of Elementary and Astroparticle Physics: Science and Technology (KSETA)," where the concluding phase of this work took place.
\bibliographystyle{utcaps_mod}
\bibliography{nugeom.bib}
\end{document}